\documentclass[fleqn,usenatbib]{mnras}

\usepackage[T1]{fontenc}

\DeclareRobustCommand{\VAN}[3]{#2}
\let\VANthebibliography\thebibliography
\def\thebibliography{\DeclareRobustCommand{\VAN}[3]{##3}\VANthebibliography}

\usepackage{graphicx}	
\usepackage{amsmath}	
\usepackage{amssymb}	
\usepackage{rotating}
\usepackage{subfig}

\usepackage{newtxtext,newtxmath}

\title[Merging neutron stars]{Predicted rates of merging neutron stars in galaxies}

\author[M. Molero et al.]{
Marta Molero,$^{1}$\thanks{E-mail: martamolero94@libero.it}
Paolo Simonetti$^{1,2}$
Francesca Matteucci,$^{1,2,3}$
Massimo della Valle$^{4}$
\\
$^{1}$Dipartimento di Fisica, Sezione di Astronomia, Università degli Studi di Trieste, Via G. B. Tiepolo 11, I-34143 Trieste, Italy\\
$^{2}$INAF, Osservatorio Astronomico di Trieste, Via G.B. Tiepolo 11, I-34143 Trieste, Italy\\
$^{3}$INFN, Trieste, Via A. Valerio 2, I-34127 Trieste, Italy\\
$^{4}$Capodimonte Astronomical Observatory, INAF-Napoli, Salita Moiariello 16, 80131-Napoli, Italy 
}

\date{Accepted XXX. Received YYY; in original form ZZZ}

\pubyear{2020}

\begin{document}
\label{firstpage}
\pagerange{\pageref{firstpage}--\pageref{lastpage}}
\maketitle

\begin{abstract}
We compute rates of merging neutron stars (MNS) in different galaxies, as well as the cosmic MNS  rate in different cosmological scenarios. Our aim is to provide predictions of kilonova rates for future observations both at low and high redshift. In the adopted galaxy models, the production of r-process elements either by MNS or core-collapse supernovae is taken into account. To compute the MNS rates, we adopt either a constant total time delay for merging (10 Myr) or a distribution function of such delays. We conclude:  i) the observed present time MNS rate in our Galaxy is well reproduced either with a constant time delay or a distribution function $\propto t^{-1}$. ii) The [Eu/Fe] vs. [Fe/H] relation can be well reproduced with only  MNS, if the time delay is short and constant. If a distribution function of delays is adopted, core-collapse supernovae are also required. iii) The present time cosmic MNS rate can be well reproduced in several cosmological scenarios. iv) Spiral galaxies are the major contributors to the cosmic MNS at all redshifts in hierarchical scenarios. In the pure luminosity evolution scenario, the spirals are the major contributors locally, whereas at high redshift ellipticals dominate. v) The predicted cosmic MNS rate well agrees with the cosmic rate of short Gamma Ray Bursts, if the distribution function of delays is adopted in a cosmological hierarchical scenario observationally derived. vi) Future observations of Kilonovae in ellipticals will allow us to disentangle among constant or a distribution of time delays as well as among different cosmological scenarios.

\end{abstract}




\section{Introduction}
\label{introduction}

The merging of two neutron stars due to gravitational wave radiation produces a strong gravitational wave (GW) signal together with its optical counterpart, the "kilonova" (powered by the decay of heavy r-process elements synthesized during the merging event) and possibly a short gamma-ray burst (SGRB). The strong connection between all these different physical phenomena with merging neutron stars (MNS) has been proved thanks to the detection of the event GW$170817$, by the Laser Interferometer Gravitational-Wave Observatory (LIGO) \citep{abbott2017a}, the first detection of a GW due to a MNS. Moreover, the observation of the kilonova AT$2017$gfo following the event, allowed us to detect the presence of heavy nuclei \citep{watson, pian, smartt}, as well as to the localize the neutron star binary system in the galaxy NGC $4993$, an early-type galaxy with an old stellar population \citep{coulter}. However, it is not to rule out the probability that the galaxy NGC $4993$ underwent a recent galactic merger \citep{ebrova}, thus the neutron stars binary system which gave rise to GW$170817$ may have come from the second accreted galaxy, which seems to be a smaller late-type galaxy.

The majority of heavy nuclei beyond the iron peak originate from neutron captures. The neutron capture on a heavy seed, such as Fe, can occur slowly or rapidly, relative to the $\beta$-decay process. Therefore, we are dealing with\textit{ s-process} or \textit{r-process} elements. The r-process elements, such as Eu, can be produced  in core-collapse supernovae (CC-SNe) \citep{woosley, Wheeler_1998} and MNS \citep{Korobkin, Rosswog_2013, hot13}.

In the context of galactic chemical evolution, it is possible to investigate the sites of production of chemical elements by using observational constraints such as the present day rate and the stellar chemical abundances.

In several chemical evolution models, the MNS  have been included as r-process element producers. In some of these models (\citealp{cescutti15}; \citealp{matteucci2014}, from now on M14) a constant gravitational time delay has been adopted for all the systems, whereas in others a more realistic delay time distribution function for such timescales (DTD) has been tested (\citealp{cote, hot}; \citealp{S19}, from now on S19). The main conclusion of M14 is that MNS cannot reproduce the [Eu/Fe] ratios in Galactic stars unless a very short and constant gravitational time delay is assumed. On the other hand, if one allows longer timescales then Eu should be produced also by CC-SNe, especially at early times.
Long time delays are also requested by the fact that the event GW170817 occurred in an early-type galaxy where star formation has stopped several Gyrs ago, and by the cosmic rate of SGRBs.



The preferred model for the  SGRBs is the merger of two compact objects (neutron stars and/or black holes), as a result of gravitational inspiral \citep{narayan}. 
Previous results (S19) have showed that the distribution of SGRBs is better reproduced with DTD $\propto t{-1}$, while too short timescales are not able to produce a good agreement with observations. Another problem with short timescales arises also from the fact that $\simeq 30\%$ of the $26$ SGRBs with classified host galaxies are found in early-type galaxies \citep{berger, davanzo} and a similar fraction has been derived by \citealp{fong} with the $36$ short GRBs detected between $2004$ and $2017$. This fact appears to be strongly in favour of long coalescence timescales for MNS. S19 also concluded that if one wants to reconcile the observed occurrence of MNS in early type galaxies with the cosmic SGRB rate and the [Eu/Fe] vs [Fe/H] in the Milky Way by assuming that only MNS produce Eu, then the DTD should contain long delay times but the fraction of binary systems giving rise to MNS should vary in time, an hypothesis that still needs to be proven.


Here we aim at computing, for the first time, the rate of MNS in galaxies of different morphological type (ellipticals, spirals and irregulars), and making predictions for future observations (e.g. LSST, VST, THESEUS). Moreover, we study the effect of MNS on the evolution of the [Eu/Fe] vs [Fe/H]. In order to do that, we use both a DTD containing long delays and a constant delay time for MNS. We compare our results of the MNS rate for a typical spiral galaxy with the Milky Way observations, such as the local rate of MNS ($\sim80^{+200}_{-60} Myr^{-1}$) derived by \citealp{kalogera2004} and the  [Eu/Fe] vs [Fe/H] pattern, and derive constraints on the main Eu producers. In particular, we compare the Eu enrichment of MNS to that of CC-SNe. Finally, we compute the cosmic MNS rate (CMNSR) in three different cosmological scenarios of galaxy formation and compare the theoretical CMNSR with that observed by LIGO/Virgo ($1540^{+3200}_{-1220}Gpc^{-3}yr^{-1}$, \citealp{abbott2017a}). and also with the SGRBs redshift distribution reconstructed by \citealp{G16}. From these comparisons we derive constrains on the origin of r-process elements as well as on the formation and evolution of galaxies of different morphological type.

This paper is organized as follows: in Section \ref{chemical evolution model} we describe the model adopted to follow the chemical evolution of galaxies of different morphological type. In Section \ref{the rate of MNS} we define the rate of MNS and describe the different DTDs for MNS adopted in this work, as well as the \citealp{greggio} DTD adopted for SNeIa. In Section \ref{simulations} we describe our simulations and present our results in terms of the predicted rate of MNS and of the predicted [Eu/Fe] vs [Fe/H] patterns in all type of galaxies. In Section \ref{cosmic} we present our analysis of the evolution of the CMNSR in three cosmological scenario and we show a comparison of the CMNSR with the redshift distribution of SGRBs. Finally, in Section \ref{conclusions} we draw our main conclusions.

\section{Chemical Evolution Model}
\label{chemical evolution model}

In this work, the evolution of galaxies of different morphological type (ellipticals, spirals and irregulars) has been studied. It is assumed that galaxies form by infall of primordial gas in a pre-existing diffuse dark matter halo. The stellar lifetimes are taken into account, thus relaxing the instantaneous recycling approximation (IRA). The model is able to follow in detail the chemical evolution of 22 elements, from H to Eu during 14 Gyr, with timesteps of 2 Myr.

The set of equations which describes the evolution of the surface mass density of the gas, $\sigma_{i}(r,t)$, in the form of the generic element $i$, are:

\begin{equation}
\begin{split}
  \dot\sigma_{i}(r,t) = & -\Psi(t)X_i(t) + X_{iA}(t)A(t)-X_{i}(t)W(t) + \dot R_{i}(t),
\end{split}
\end{equation}
\\
where $X_i(t)=\sigma_i(t)/\sigma_{gas}(t)$ is the abundance by mass of the element $i$ at the time $t$.

The terms on the right-hand side of the equation are the following:

\begin{itemize}
    \item The first term is the rate at which chemical elements are subtracted by the interstellar medium (ISM) to be included in stars. $\Psi(t)$ is the star formation rate (SFR), which represents how many solar mass of gas is turned into stars per unit time e per unit area. Here we parametrized the SFR adopting the Schmidt-Kennicutt law (\citealp{schmidt, kennicutt}):
    
    \begin{equation}
    \Psi(t)=\nu\sigma^{k}_{gas},
    \end{equation}
    \\
    with  $k=1$ and the star formation efficiency $\nu$ which varies according to the morphological type of galaxy (see Table 1);
    \end{itemize}
    
    \begin{itemize}
    \item The second term is the rate at which chemical elements are accreted through infall of gas. $X_{i,A}(t)$ describes the chemical abundance of the element i of the infalling gas, assumed to be primordial, and $A(t)$ represents the gas accretions rate, here described by an exponential law:
    
    \begin{equation}
    A(t)=ae^{-t/\tau_{inf}},
    \end{equation}
    \\
    where $a$ is a normalisation constant constrained to reproduce the present time total total surface mass density and $\tau_{inf}$ is the infall timescale, defined as the time at which half of the total mass of the galaxy has assembled, it depends on the morphological type of the galaxy, increasing from ellipticals to spirals and irregulars.
    \end{itemize}
    
    \begin{itemize}
    \item The third term is the rate at which chemical elements are lost through galactic winds. $W(t)$ is the wind rate, assumed to be proportional to the SFR:
    
    \begin{equation}
    W(t)=-\omega\Psi(t),
    \end{equation}
    \\
    where $\omega$ is a free parameter measuring the efficiency of the wind and which can vary according to the morphological type of the galaxy. Here $\omega$ has been assumed to be equal for all elements and chosen in order to reproduce the characteristic features of the galaxy. Also, the galactic winds are assumed to develop when the thermal energy on the gas (heated by the supernovae explosions) exceeds its binding energy (\citealp{bradamante}).
    \end{itemize}
    
    \begin{itemize}
    \item The last term represents the rate of restitution of matter into the ISM from all the stars dying at the time $t$. 
    It consists of the SN rates (II, Ia, Ib, Ic) plus the MNS rate convolved with the stellar nucleosynthesis, as well as of all the contributions from low and intermediate mass stars (LIMS). In other words, this term contains the rates at which each element is restored by different stellar sources. It depends also on the initial mass function  (IMF), $\varphi(m)$. In particular, we have adopted here a \citep{salpeter} IMF for all galaxy types:
    
    
    \begin{equation}
    \varphi(t)=0.17m^{-(1+1.35)},
    \end{equation}
    \\
    normalised to unity between $0.1$ and $100 M_\odot$. 
\end{itemize}

For all the stars sufficiently massive to die in a Hubble time, the following stellar yields have been adopted:

\begin{itemize}
    \item For LIMS with mass lower than $6 M_{\odot}$ we used the yields by \citealp{karakas};
    \item For super-AGB (SAGB) stars and e-capture SNe with masses between $6 M_{\odot}$ and $10 M_{\odot}$ we used the yields by \citealp{doherty};
    \item For massive stars that explode as CC-SNe we used the yields by \citealp{nomoto};
    \item For Type Ia SNe we used the yields by \citealp{iwamoto}. Their rate has been computed by convolving the adopted SFR with the \citealp{greggio} delay time distribution (DTD) for wide double degenerate SNeIa.
\end{itemize}

For what concerns the yields of Eu from massive stars, they are the same adopted by S19, which are a modified version of those found in \citealp{argast} (their model SN$2050$) and also used by M14, in particular:

\begin{itemize}
    \item For stars in the $20-23$ $M_{\odot}$ mass range, a constant yield of $3.8 \times 10^{-8}$ $M_{\odot}$ of Eu has been used;
    \item A decreasing yield from $3.8 \times 10^{-8}$ $M_{\odot}$ of Eu for a $23$ $M_{\odot}$ star to $1.7 \times 10^{-9}$ $M_{\odot}$ of Eu for a $50$ $M_{\odot}$ star has been used.
\end{itemize}
    
{It should be noted that we assume that Eu is produced only by a fraction of CC-SNe, in particular from those in the range 20-50$M_{\odot}$. Therefore, the rate of CC-SNe producing Eu is only a fraction (a factor of $\sim 5$ less) of the total CC-SN rate, which is related to the entire range of massive stars from 8 to 100$M_{\odot}$.}

The yield of Eu from MNS, instead, is described following the theoretical calculations of \citealp{Korobkin}, who estimate the production of Eu from each event to be in the range of $10^{-7}-10^{-5}$ $M_{\odot}$.

It is worth noting that there is a degeneracy between the Eu yields and the rates of its progenitors (MNS and CC-SNe). In fact, M14 first showed that it is possible to reproduce the [Eu/Fe] vs [Fe/H] pattern in the Galaxy only with MNS by adopting a specific yield from a merging event. However, if one allows CC-SNe to produce Eu as well, then the yield of Eu from MNS should be lower. Again, if only the rare class of magneto-rotationally driven SNe are included (see \citealp{cescutti15}), the yield of Eu from MNS changes again.

\begin{figure}
\begin{center}
  \includegraphics[width=0.9\linewidth]{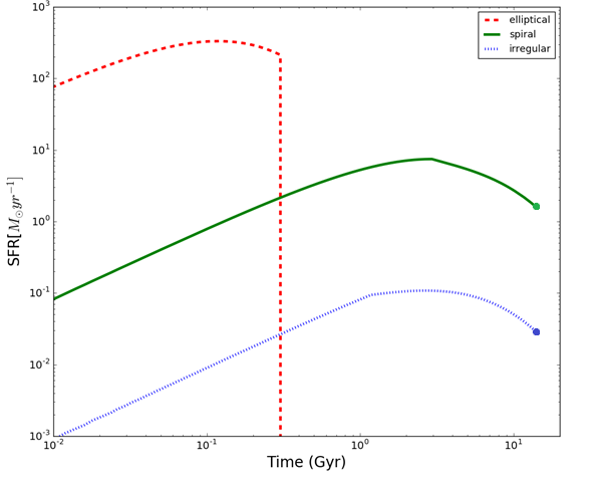}
  \caption{Predicted SFRs for galaxies of different morphological type as a function of time. Red dashed, green solid and blue dotted lines represent elliptical, spiral and irregular galaxies, respectively. Present time data are taken from \protect\citealp{chomiuk} (green circle) and \protect\citealp{rubele} (blue circle) for the Galaxy and for the Small Magellanic Cloud, respectively.}
  \label{fig: SFRdm}
 \end{center}
\end{figure}

The input parameters adopted for the different galaxies are reported in Table \ref{tab: Input parameters DM}, where we specified in the first column the type of galaxy, in the second column the infall mass, in the third column the star formation efficiency, in the fourth column the infall timescale, in the fifth column the effective radius and in the sixth column the wind parameter. As it is shown in Figure \ref{fig: SFRdm}, the parameters for spiral and irregular galaxies have been fine tuned in order to reproduce the measured present day SFR in the solar neighbourhood \citep{chomiuk} and in the Small Magellanic Cloud \citep{rubele}, respectively. For elliptical galaxies, instead, the parameters adopted trace the typical behaviour of an elliptical galaxy with a quenching of the star formation, determined by the action of the galactic winds, after an initial and very intense burst (see \citealp{pipinomatteucci}). The SFR of ellipticals is higher than the one of irregulars and spirals, according to the downsizing scenario for which larger galaxies have higher star formation efficiency \citep{matteucci1994}. On the other hand, the simulated spiral galaxy is characterized by a continuous SFR which is higher than the one of irregulars.

\begin{table*}
\centering
\hspace{-0.5 cm}
\caption{\label{tab: Input parameters DM}Parameters used for the chemical evolution models of spiral, irregular and elliptical galaxies. In the first column it is reported the morphological type of the galaxy, in the second, third, fourth and fifth column the infall mass $M_{infall}$, the star formation efficiency $\nu$, the infall timescale $\tau_{infall}$, the effective radius $R_{eff}$, and the wind parameter $\omega_{i}$, respectively.}
\begin{tabular}{lccccr}
\hline
\hline
  Type & $M_{infall}$ $(M_{\odot})$ & $\nu$ $(Gyr^{-1})$ & $\tau_{infall}$ $(Gyr)$ & $R_{eff}$ $(pc)$ & $\omega_{i}$ \\
\hline
  Spiral & $5.0 \times 10^{10}$ & $1$ & $7$ & $3.5 \times$ $10^3$ & $0.2$\\
  Irregular & $5.5 \times 10^{8}$ & $0.1$ & $7$ & $1 \times$ $10^3$ & $0.5$\\
  Elliptical & $5.0 \times 10^{11}$ & $17$ & $0.2$ & $7 \times$ $10^3$ & $10$\\
\hline
\hline
\end{tabular}%
\end{table*}

\begin{figure*}
\begin{center}
 \subfloat[]{\includegraphics[width=1\columnwidth]{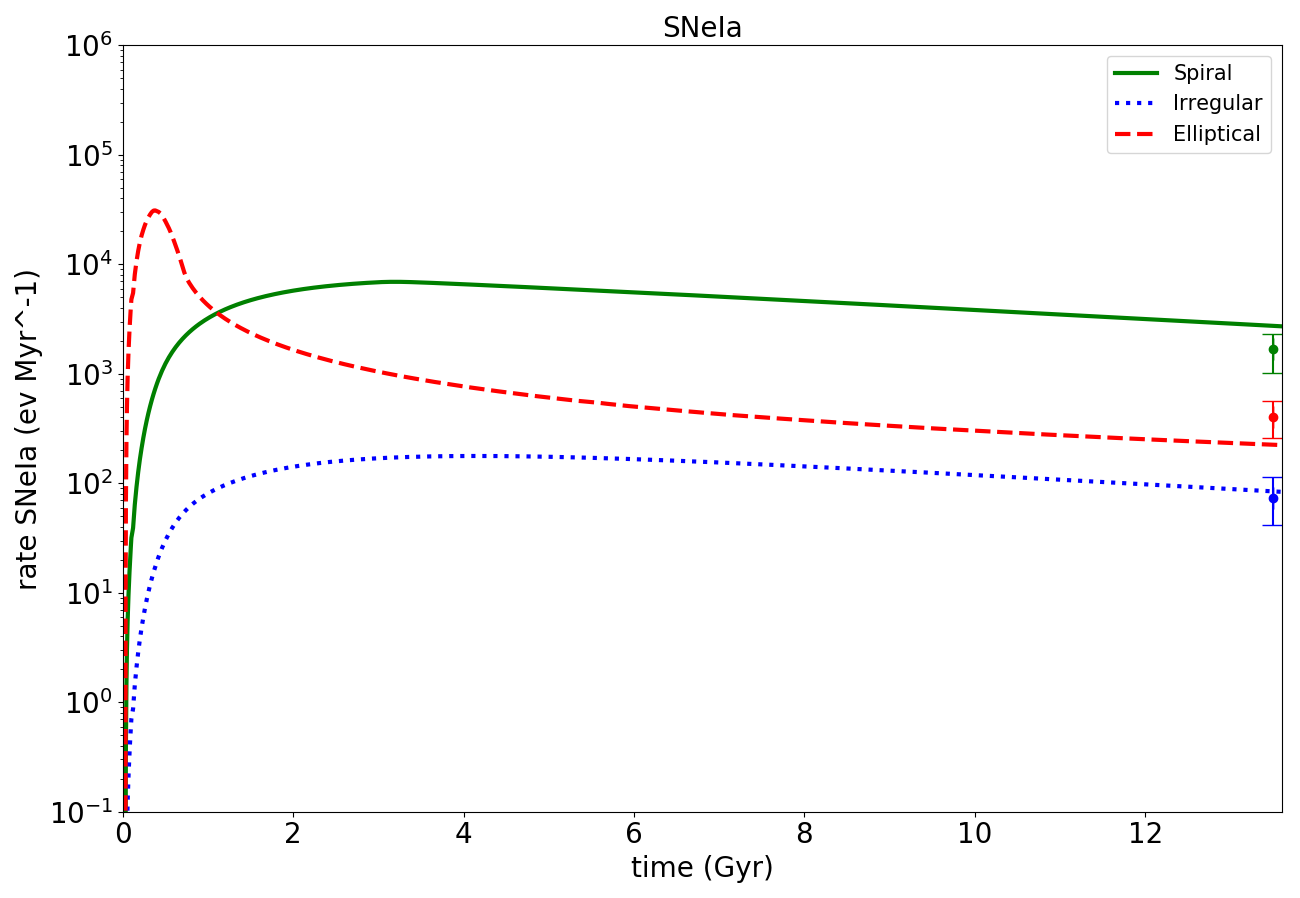}\label{fig:a}}
 \hfill
 \subfloat[]{\includegraphics[width=1\columnwidth]{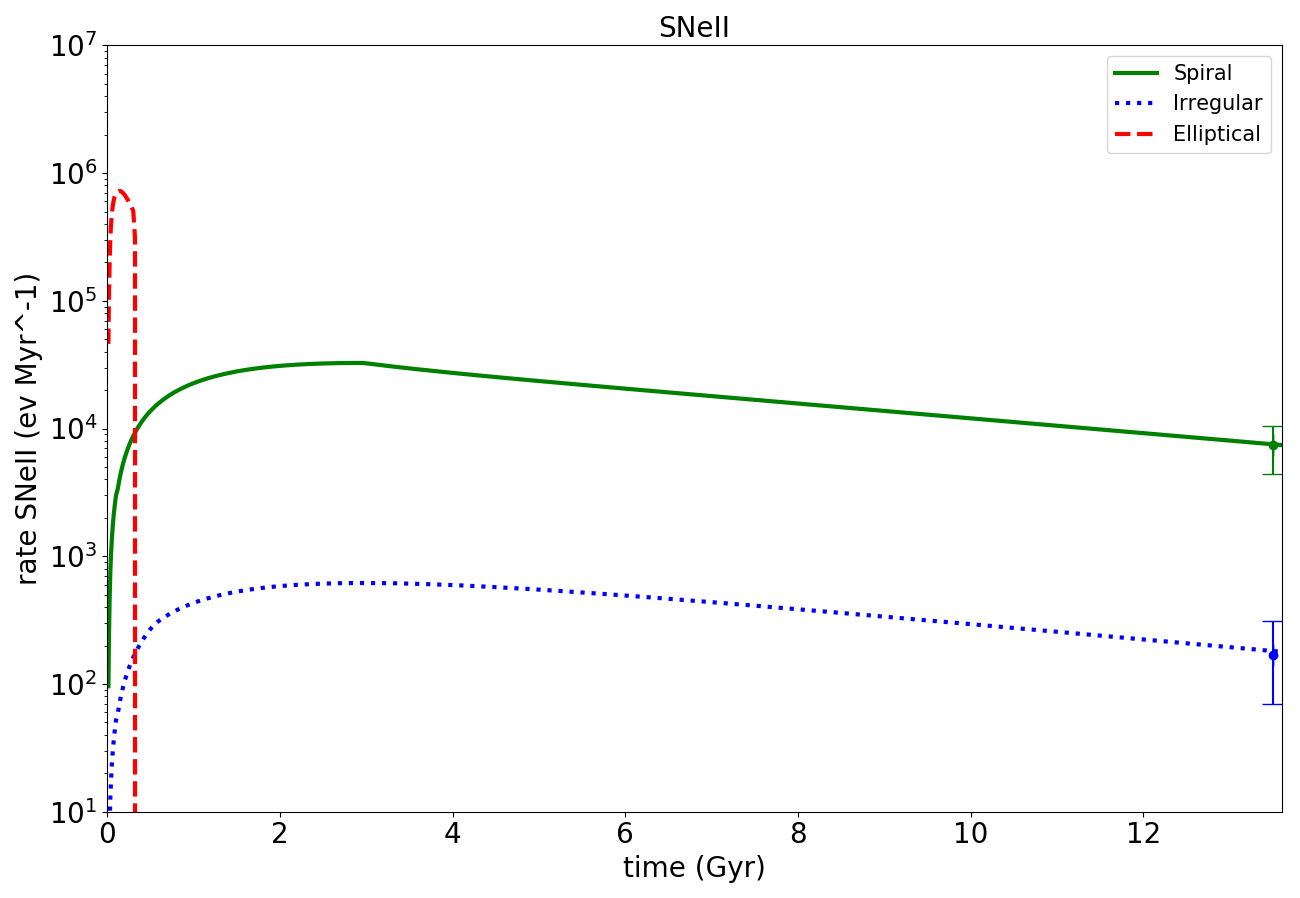}\label{fig:b}}%
 \caption{Predicted Type Ia SN (panel (a)) and CC-SN (panel (b)) rates for spiral, elliptical and irregular galaxies. Present time rates are indicated and are from \protect\citealp{mannucci05}}%
 \label{fig: SNrate}%
\end{center}
\end{figure*}

In order to show that our galaxy models reproduce the majority of the observed features, we show in Figure \ref{fig: SNrate} the Type Ia SN and CC-SN rates for the three morphological types of galaxies, in units of $events\times Myr^{-1}$. Together with our predicted rates are reported the present time values of the observed SN rates in each galaxy type. In the case of elliptical galaxies, the present time CC-SN rate is null since these galaxies do not show any trace of young stars. As one can see from the Figures, we reproduce quite well the present time rates for all galaxies.

\section{The Rate of Merging Neutron Stars}
\label{the rate of MNS}

In the context of chemical evolution of galaxies, the timescales for the production of each element are of great importance. As already described in section \ref{introduction}, the two main sites that have been identified for the production of r-process elements are CC-SNe and MNS. In the case of CC-SNe, the global timescale of production of r-process elements is given by the nuclear lifetime of the massive stars. On the other hand, for what concerns MNS, the timescale is given by the sum of the progenitor lifetime and the delay due to gravitational wave radiation, which depends on the initial separation of the binary system and on the masses of the two components.

In this work, the rate of MNS in galaxies of different morphological type has been computed using two different approaches. The first one consisted in assuming a constant delay time between the formation of the neutron stars binary system and the merging event, while the second one consisted in assuming a probability distribution of delay times. The delay time distribution (DTD) for MNS represents the distribution of the coalescence times of MNS formed in an instantaneous burst of star formation, namely, a single stellar population of unitary mass. It gives the probability of the merging event to occur at a given time $t$ from the formation of the neutron stars binary system progenitors. Then, the MNS rate $R_{MNS}(t)$ is simply obtained as the convolution of a given DTD with a given star formation rate (SFR):

\begin{equation}
 R_{MNS}(t)=k_{\alpha} \int_{\tau_{i}}^{min(t,\tau_{x})} \! \alpha_{MNS}(\tau)\Psi(t-\tau)f_{MNS}(\tau) \, \mathrm{d}\tau,
\label{eq:ratemns}
\end{equation}
\\
where $\Psi$ is the SFR and $f_{MNS}$ is the DTD. $\alpha_{MNS}$ is the fraction stars in the correct mass range which can give rise to a double neutron star merging event, and in principle it can vary with time, but here it is assumed to be constant. $\tau$ is the total delay time defined in the range $(\tau_i,\tau_x)$, so that:

\begin{equation}
    \int_{\tau_{i}}^{\tau_{x}} \! f_{MNS}(\tau) \, \mathrm{d}\tau=1,
\end{equation}
\\
where $\tau_i$ is the minimum delay time of occurrence for MNS (here fixed at 10 Myr) and $\tau_x$ is the maximum delay time which can be larger than a Hubble time. Finally, $k_\alpha$ is the number of neutron stars progenitors per unit mass in a stellar generation and it depends on the IMF, in particular:

\begin{equation}
    k_\alpha=\int_{m_{m}}^{m_M} \! \varphi(m) \, \mathrm{d}m,
\end{equation}
\\
where $m_m=9 M_\odot$ and $m_M=50 M_\odot$ are respectively the progenitor minimum and maximum mass to produce a neutron star. In our case, since we adopted a Salpeter IMF, we have $k_\alpha=5.87\times 10^{-3}$.

This formulation has the advantage to use any function to describe the DTD and also to test the empirically derived ones.

\begin{figure}
\begin{center}
  \includegraphics[width=1\linewidth]{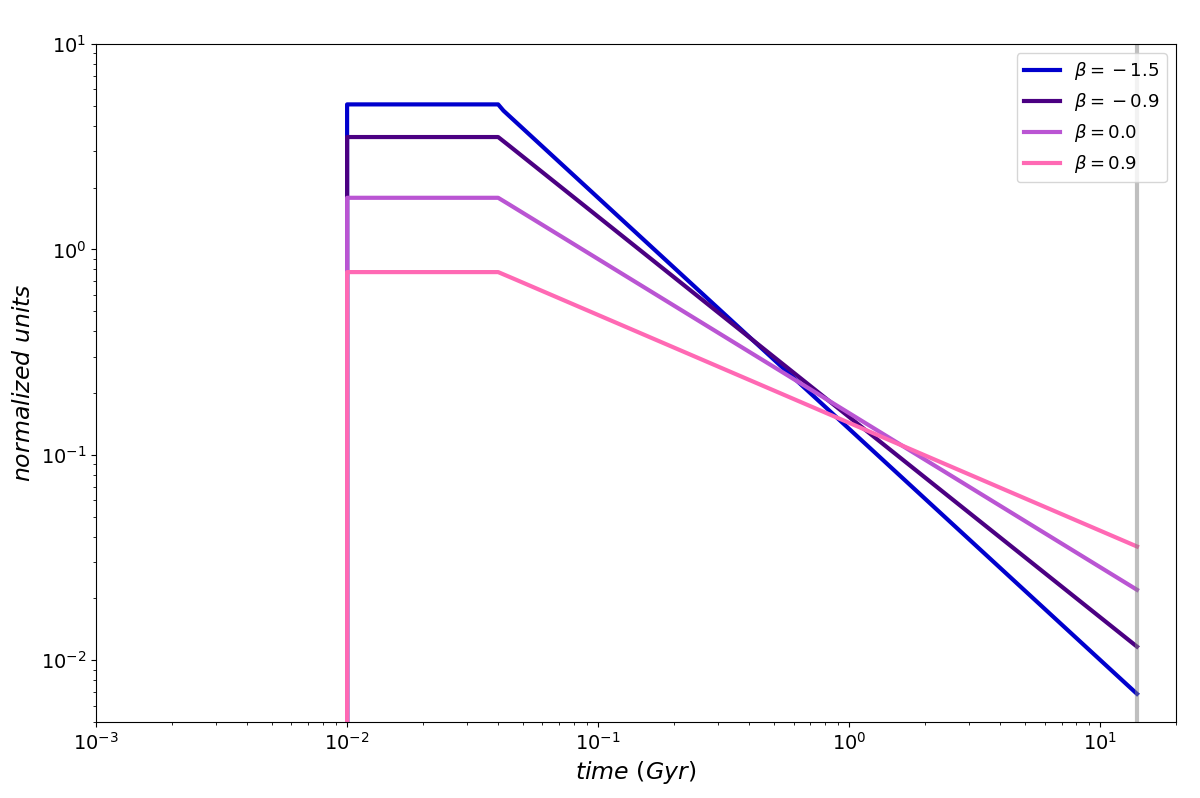}
  \caption{Delay time distributions for MNS computed for the four different values of $\beta$ used in this thesis. Lower values of $\beta$ represent neutron stars binary systems with shorter initial separation. A normalization factor of $1$ has been assumed.}
  \label{fig: DTD}
 \end{center}
\end{figure}

For MNS the delay time is given by the sum of the nuclear lifetime and the delay due to the emission of gravitational waves, namely the gravitational time delay. In this work we adopted the following distribution of the total (nuclear plus gravitational) delay time, derived by S19:

\begin{equation}
    f(\tau)\propto 
    \begin{cases}
    \ 0 \qquad if \qquad \tau < 10 Myr \\
    \ p_1 \qquad if \qquad  10 < \tau < 40 Myr \\
    \ p_2\tau^{0.25\beta -0.75}(M_{m}^{0.75(\beta + 2.33)}-M_{M}^{0.75(\beta + 2.33)}) \\ 
    \ if \qquad  40 Myr < \tau < 13.7 Gyr
    \end{cases}
\label{eq: DTD}
\end{equation}
\\
where $\beta$ is the parameter which characterizes the shape of the initial separation function. In particular, here we tested four different values of $\beta$, which are reported in Table \ref{tab: p1p2} together with the adopted values for $p_1$ and $p_2$ which must be chosen in order to obtain a continuous and normalized function. $M_m$ and $M_M$ are the minimum and maximum total mass of the system, respectively. In Figure \ref{fig: DTD} we show the DTDs tested in this work. As one can see, the first portion of the distribution ends with the formation of the first double neutron star system. 10 Myr is in fact the nuclear lifetime of a typical massive star. The second portion refers to systems which merge soon after the formation of the double neutron star systems. This portion of the distribution is described by a flat plateau, up to the lifetime of the minimum mass progenitor of a neutron star. The third part of the distribution is the distribution of the gravitational delay times and attains to those systems for which the time delay is dominated by gravitational radiation.

\begin{table}
\hspace{-0.5 cm}
\caption{\label{tab: p1p2}Values of $p_1$ and $p_2$ for different $\beta$ chosen in order to obtain a continuous and normalized DTD function, as expressed by eq. \ref{eq: DTD}.}
\centering
\begin{tabular}{ccc}
\hline
\hline
    $\beta$ & $p_1$ & $p_2$ \\
\hline
    $-1.5$ & $5.076$ & $0.159$ \\
    $-0.9$ & $3.521$ & $0.065$ \\
    $0.0$ & $1.783$ & $0.020$ \\
    $0.9$ & $0.775$ & $0.006$ \\ 
\hline
\hline
\end{tabular}
\end{table}

For a complete discussion about the derivation of the DTD for MNS we suggest to see S19.

\subsection{Delay Time Distribution for SNeIa}
\label{DTD SNEIA}

Since one of our purposes is studying the [Eu/Fe] vs [Fe/H] patterns in galaxies of different morphological type, we need also to compute the Fe evolution and to describe how Fe production from SNeIa (the most important contributors to this element) has been implemented in our models. 

Adopting the same formulation used for the rate of MNS, the rate of SNeIa has been computed as:

\begin{equation}
 R_{Ia}(t)=k_{\alpha, Ia} \int_{\tau_{i}}^{min(t,\tau_{x})} \! \alpha_{Ia}(\tau)\Psi(t-\tau)f_{Ia}(\tau) \, \mathrm{d}\tau,
\label{eq:ratemns}
\end{equation}
\\
where $\alpha_{Ia}$ is the fraction of binary systems giving rise to SNeIa. This quantity can vary with time, but here we assume it to be constant. $f_{Ia}$ is the DTD of SNeIa which, in analogy with the DTD for MNS, represents the distribution of the explosion times from an instantaneous burst of star formation of unitary mass. It must be normalized to $1$ in the allowed range for the delay time $\tau$:

\begin{equation}
    \int_{\tau_{i}}^{\tau_{x}} \! f_{Ia}(\tau) \, \mathrm{d}\tau=1,
\end{equation}
\\
where $\tau_i$ is the minimum total delay time of occurrence of type Ia SNe, here fixed at 40 Myr (corresponding to the lifetime of a $8 M_\odot$ star) and $\tau_x$ is the maximum total delay time which can be larger than a Hubble time, according to the chosen progenitor model. In this work we adopted for SNeIa the DTD suggested by \citealp{greggio} for the wide double degenerate (DD) scenario:

\begin{equation}
    f_{Ia} \propto \int_{\tau_{n,i}}^{min(\tau_{n,x}, \tau)} \! n(\tau_n)S(\tau,\tau_n), \mathrm{d}\tau_n,
\end{equation}
\\
where $\tau_{n,i}$ and $\tau_{n,x}$ are the nuclear lifetimes of the most and least massive secondary in the SNeIa  progenitors binary systems, respectively. $n(\tau_n)$ is the distribution function of the nuclear delays of the SNIa progenitors and

\begin{equation}
    S(\tau,\tau_n)=
    \begin{cases}
    \ (M_{m}^{0.75\beta_{Ia} + 1.75}-M_{M}^{0.75\beta_{Ia} + 1.75})(\tau-\tau_n)^{-0.75+0.25\beta_{Ia}} \\
    \ if \qquad \tau_n \le \tau-\tau_{gw,i} \\
    \ 0 \qquad if \qquad \tau_n \ge \tau-\tau_{gw,i}
    \end{cases}
\end{equation}
where $\tau_{gw,i}$ is the minimum gravitational time delay. We adopted a distribution characterized by an initial separation function with exponent equal to $\beta_{Ia}=-0.9$ and a maximum nuclear delay time $\tau_{n,x}=0.4$ Gyr. 

Finally, $k_{\alpha, Ia}$ is the number of stars per unit mass in a stellar generation, in particular:

\begin{equation}
    k_{\alpha, Ia}=\int_{m_{L}}^{m_U} \! \varphi(m) \, \mathrm{d}m,
\end{equation}
\\
where $m_L=0.1 M_\odot$ and $m_U=100 M_\odot$. In our case, we have $k_{\alpha, Ia}=2.83$.

\begin{figure}
\begin{center}
  \includegraphics[width=1\linewidth]{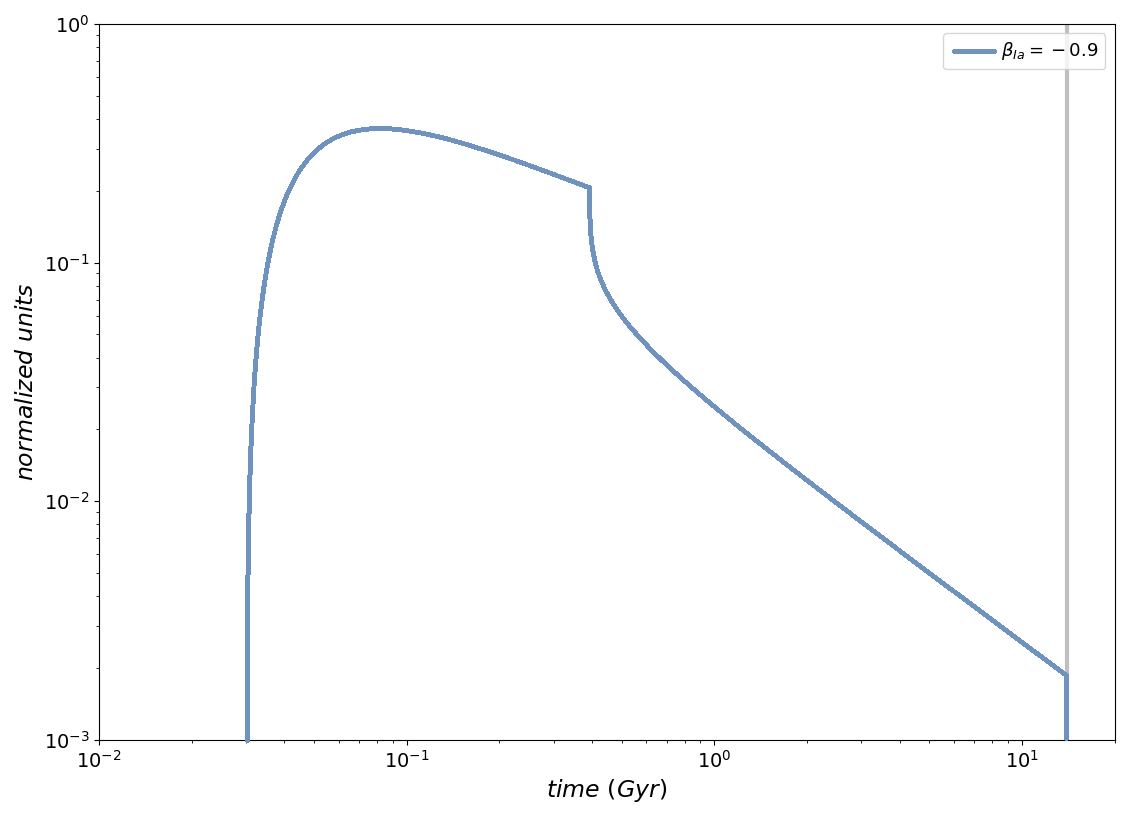}
  \caption{Delay time distribution for SNeIa obtained from the \protect\citealp{greggio} wide DD model with $\beta_{Ia}=-0.9$.}
  \label{fig: dtdsneia}
 \end{center}
\end{figure}

The adopted distribution is shown in Figure \ref{fig: dtdsneia}.

\section{Simulations for Galaxies of Different Morphological Type}
\label{simulations}

In order to predict the rate of MNS and the [Eu/Fe] vs [Fe/H] pattern in galaxies of different morphological type (Ellipticals, Spirals and Irregulars) we run several chemical evolution models where we need to specify the following parameters:

\begin{enumerate}
    \item The DTD of MNS;
    
    \item The fraction of neutron stars in binaries which produce a MNS, $\alpha_{MNS}$;
    
    \item The production of Eu by CC-SNe;
    
    \item The Eu produced per merging event.
\end{enumerate}

In particular:

\begin{itemize}
    \item We tested four different DTDs derived by S19 corresponding to four different values of $\beta$ ($-1.5$, $-0.9$, $0.0$, $0.9$) as described in Section \ref{the rate of MNS}, as well as a constant total delay time (which includes both the nuclear lifetime and the time necessary to merge) for all the neutron stars binary systems, equal to 10 Myr;
    \item The parameter $\alpha_{MNS}$ has been fixed in order to reproduce, for spiral galaxies, the MNS rate in the Milky Way as suggested by \citealp{kalogera2004}, $\sim 80^{+200}_{-60} Myr^{-1}$;
    \item For half of the models we considered the Eu to be co-produced by CC-SNe and MNS, while for the other half we assumed that MNS were the only Eu producers in the simulated galaxy;
    \item The yield of Eu per merging event has been fixed in order to reproduce, for spiral galaxies, the solar absolute abundance of Eu as derived by \citealp{lodders}. In particular, we compared the ISM Eu abundance at 9 Gyr since the beginning of the star formation, with the solar abundance.
\end{itemize}

On the other hand, the parameters left constant are:

\begin{enumerate}
    \item The progenitor mass range for MNS, between $9$ and $50$ $M_\odot$;
    \item The CC-SNe mass range, between $20$ and $50$ $M_\odot$, with prescriptions for the yield of Eu, as described in Section \ref{chemical evolution model};
    \item Prescriptions about IMF, SFR and yield of elements other than Eu, which are also specified in Section \ref{chemical evolution model};
    \item The DTD for SNeIa, as described in Section \ref{DTD SNEIA}.
\end{enumerate}

The different models that we run are reported in Table \ref{tab: results galaxies CC} and in Table \ref{tab: results galaxies noCC}, together with their predictions. For both Tables, in the first column it is specified the name of the model; in the second column is reported the type of simulated galaxy; in the third column it is specified if CC-SNe contributed to the Eu production; in the fourth and fifth columns are reported the adopted DTD for SNeIa and the occurrence probability $\alpha_{Ia}$, respectively; in the sixth and seventh columns are shown the DTD for MNS and their occurrence probability $\alpha_{MNS}$, respectively; in the eighth column it is reported the yield of Eu per merging event and in the last column the predicted MNS rate. 

\begin{table*}
\hspace{-0.5 cm}
\caption{\label{tab: results galaxies CC} All the models for galaxies of different morphological type with Eu production from CC-SNe. In the $1^{st}$ column the name of the model, in the $2^{nd}$ column the galaxy type, in the $3^{rd}$ column the production of Eu from CC-SNe, in the $4^{th}$ column the DTD used for SNeIa, in the $5^{th}$ column the occurrence probability of SNeIa ($\alpha_{Ia}$), which has been tuned to obtain the estimated current rate of SNeIa, in the $6^{th}$ column the DTD used for MNS, in the $7^{th}$ column the occurrence probabilities of MNS ($\alpha_{MNS}$) which has been tuned in order to obtain for spiral galaxies the estimated rate of MNS in the Milky Way, in the $8^{th}$ column the yield of Eu from MNS and in the $9^{th}$ column the predicted rates of MNS.}
\centering
\begin{tabular}{lcccccccr}
\hline
\hline
Model & Galaxy & Eu from & SNeIa & $\alpha_{Ia}$ & MNS & $\alpha_{MNS}$ & MNS Eu yield & MNS rate \\
 & Type & CC-SNe & DTD & ($\times 10^{-3}$) & DTD & $(\times 10^{-2})$ & $(\times 10^{-6}M_{\odot})$ & $(Myr^{-1})$ \\
\hline
 1Sa & Spiral & yes & $\beta_a=-0.9$ & 3.29 & $\beta=-1.5$ & $5.58$ & $0.5$ & 72 \\
 2Sa & Spiral & yes & $\beta_a=-0.9$ & 3.29 & $\beta=-0.9$ & $5.42$ & $0.5$ & 80 \\
 3Sa & Spiral & yes & $\beta_a=-0.9$ & 3.29 & $\beta=0.0$ & $5.21$ & $0.5$ & 92 \\
 4Sa & Spiral & yes & $\beta_a=-0.9$ & 3.29 & $\beta=0.9$ & $5.06$ & $0.5$ & 105 \\
 5Sa & Spiral & yes & $\beta_a=-0.9$ & 3.29 & Constant 10 Myr & $6.15$ & $0.5$ & 77 \\
 1Ia & Irregular & yes & $\beta_a=-0.9$ & 4.29 & $\beta=-0.9$ & $5.58$ & $0.5$ & 12 \\
 2Ia & Irregular & yes & $\beta_a=-0.9$ & 4.29 & Constant 10 Myr & $6.15$ & $0.5$ & 13 \\
 1Ea & Elliptical & yes & $\beta_a=-0.9$ & 5.05 & $\beta=-0.9$ & $5.58$ & $0.5$ & 15 \\
 2Ea & Elliptical & yes & $\beta_a=-0.9$ & 5.05 & Constant 10 Myr & $6.15$ & $0.5$ & 0 \\
\hline
\hline
\end{tabular}
\end{table*}

\begin{table*}
\hspace{-0.5 cm}
\caption{\label{tab: results galaxies noCC} All the models for galaxies of different morphological type with no Eu production from CC-SNe. In the $1^{st}$ column the name of the model, in the $2^{nd}$ column the galaxy type, in the $3^{rd}$ column the production of Eu from CC-SNe, in the $4^{th}$ column the DTD used for SNeIa, in the $5^{th}$ column the occurrence probability of SNeIa ($\alpha_{Ia}$), which has been tuned to obtain the estimated current rate of SNeIa, in the $6^{th}$ column the DTD used for MNS, in the $7^{th}$ column the occurrence probabilities of MNS ($\alpha_{MNS}$) which has been tuned in order to obtain for spiral galaxies the estimated rate of MNS in the Milky Way, in the $8^{th}$ column the yield of Eu from MNS and in the $9^{th}$ column the predicted rates of MNS.}
\centering
\begin{tabular}{lcccccccr}
\hline
\hline
Model & Galaxy & Eu from & SNeIa & $\alpha_{Ia}$ & MNS & $\alpha_{MNS}$ & MNS Eu yield & MNS rate \\
 & Type & CC-SNe & DTD & ($\times 10^{-3}$) & DTD & $(\times 10^{-2})$ & $(\times 10^{-6}M_{\odot})$ & $(Myr^{-1})$ \\
\hline
 1Sb & Spiral & no & $\beta_a=-0.9$ & 3.29 & $\beta=-1.5$ & $5.58$ & $2.0$ & 72 \\
 2Sb & Spiral & no & $\beta_a=-0.9$ & 3.29 & $\beta=-0.9$ & $5.42$ & $2.0$ & 80 \\
 3Sb & Spiral & no & $\beta_a=-0.9$ & 3.29 & $\beta=0.0$ & $5.21$ & $2.0$ & 92 \\
 4Sb & Spiral & no & $\beta_a=-0.9$ & 3.29 & $\beta=0.9$ & $5.06$ & $2.0$ & 105 \\
 5Sb & Spiral & no & $\beta_a=-0.9$ & 3.29 & Constant 10 Myr & $6.15$ & $2.0$ & 77 \\
 1Ib & Irregular & no & $\beta_a=-0.9$ & 4.29 & $\beta=-0.9$ & $5.58$ & $2.0$ & 12 \\
 2Ib & Irregular & no & $\beta_a=-0.9$ & 4.29 & Constant 10 Myr  & $6.15$ & $2.0$ & 13 \\
 1Eb & Elliptical & no & $\beta_a=-0.9$ & 5.05 & $\beta=-0.9$ & $5.58$ & $2.0$ & 15 \\
 2Eb & Elliptical & no & $\beta_a=-0.9$ & 5.05 & Constant 10 Myr  & $6.15$ & $2.0$ & 0 \\
\hline
\hline
\end{tabular}
\end{table*}

\subsection{Predicted MNS rate}
\label{section: MNSrate}

\begin{figure}
\begin{center}
  \includegraphics[width=1\linewidth]{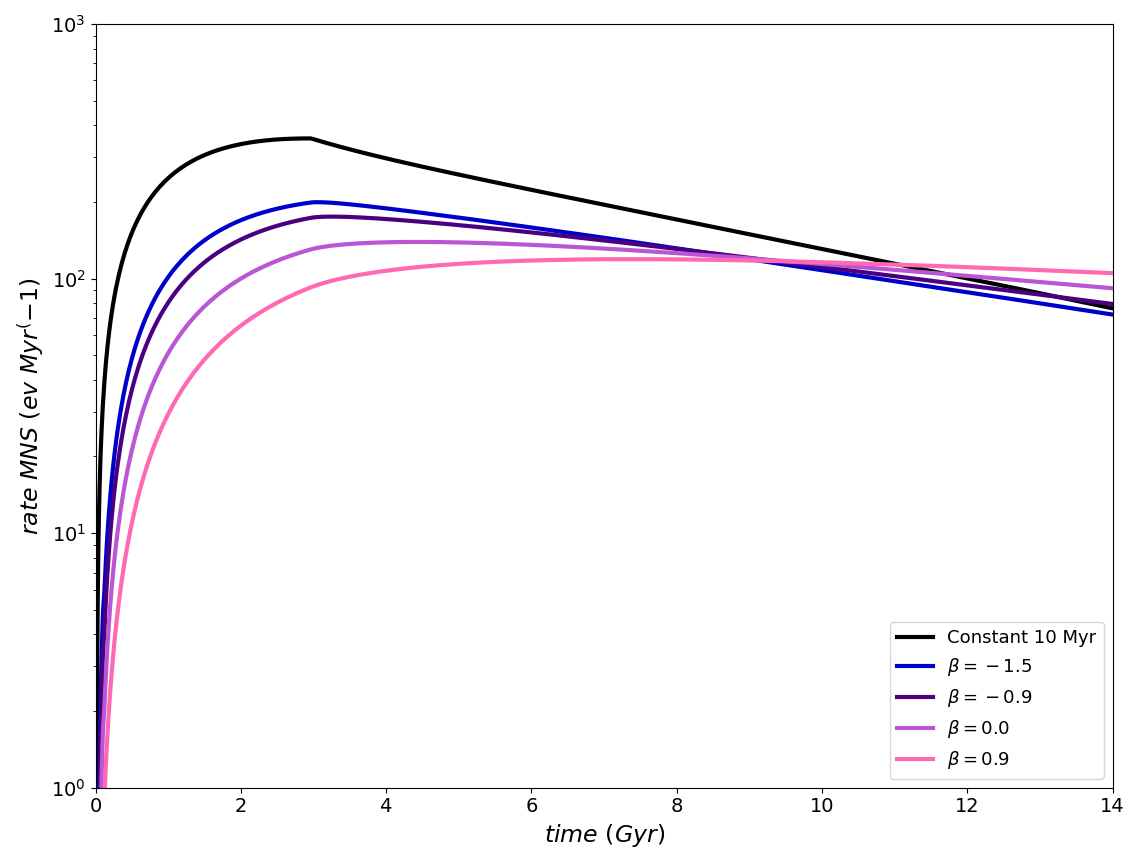}
  \caption{Predicted rate of MNS in spiral galaxies as a function of time for the four different DTDs derived by S19 corresponding to four different values of $\beta$ ($-1.5$, $-0.9$, $0.0$, $0.9$) and for a constant total delay time equal to 10 Myr.}
  \label{fig: ratespiral}
 \end{center}
\end{figure}

First, we show the results of our simulations for spiral galaxies. The predicted rate of MNS as a function of time can be seen in Figure \ref{fig: ratespiral}, where we show the rate of MNS both in the case of a DTD and in the case of a constant total delay time of 10 Myr for all neutron star binary systems. As it is clear from Figure 4, the constant and short time delay predicts a higher MNS rate at early times, relative to the cases adopting DTDs including also long delay times. On the other hand, no much difference is predicted for the present time MNS rate by all the studied cases. This is due to the fact that a typical spiral suffers continuous star formation until the present time. A large difference in the present time value of the MNS rate is instead predicted for ellipticals (see later). As it is possible to see from Tables \ref{tab: results galaxies CC} and \ref{tab: results galaxies noCC}, all of our simulations for spiral galaxies give us results consistent with the observed MNS rate for the Milky Way. More precisely, the rate is best represented by DTDs with lower values of $\beta$, so by bottom heavy distributions, in particular the one corresponding to $\beta=-0.9$. We also remind that lower values of $\beta$ imply systems with small initial separations. Also the case of a constant total delay time of 10 Myr appears to be a good candidate to represent the MNS rate. In those two cases, the occurrence probability of MNS ($\alpha_{MNS}$) is found to be $5.42\%$ and $6.15\%$, respectively. 

\begin{figure*}
\begin{center}
 \subfloat[]{\includegraphics[width=1\columnwidth]{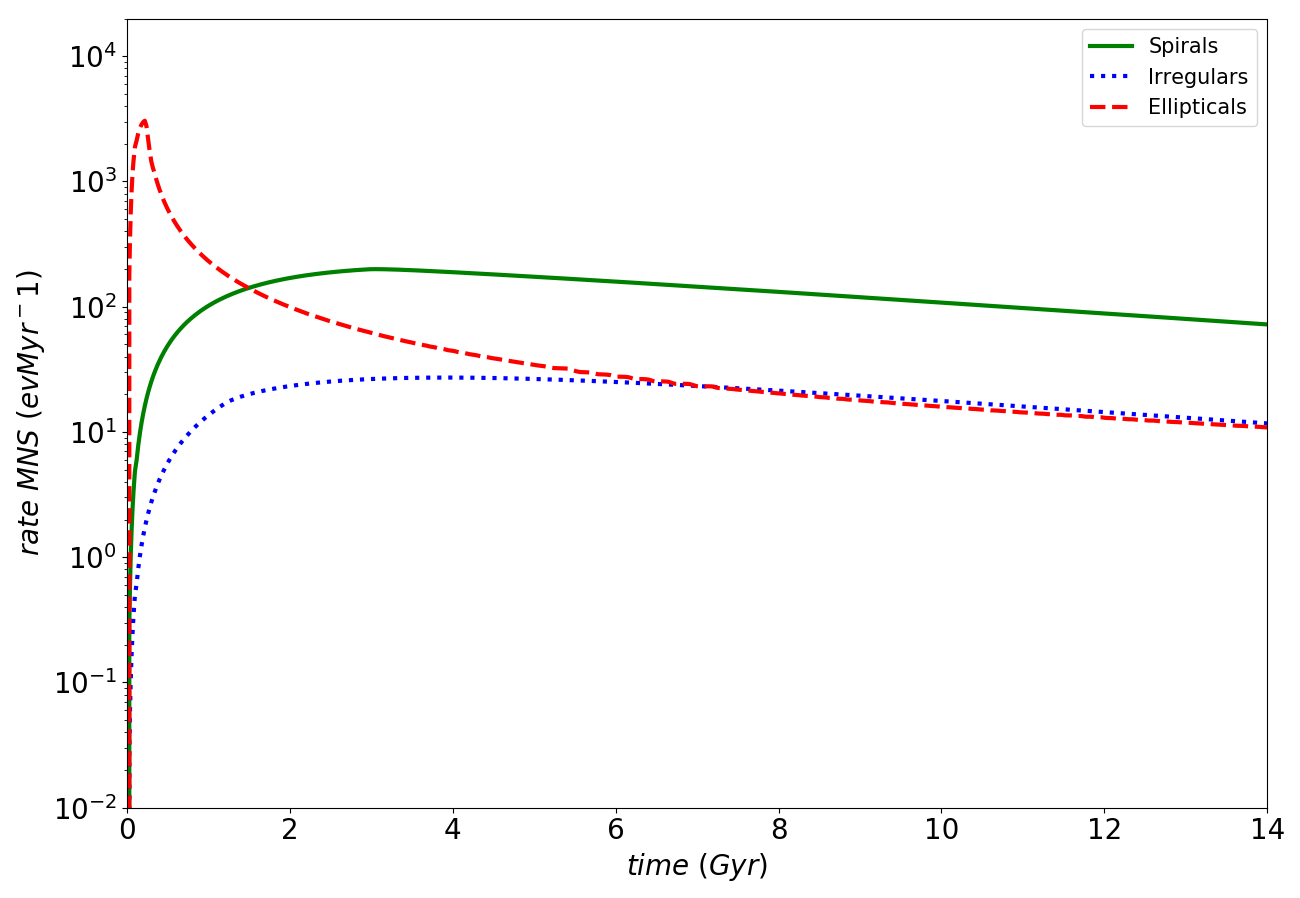}\label{fig:a}}
 \hfill
 \subfloat[]{\includegraphics[width=1\columnwidth]{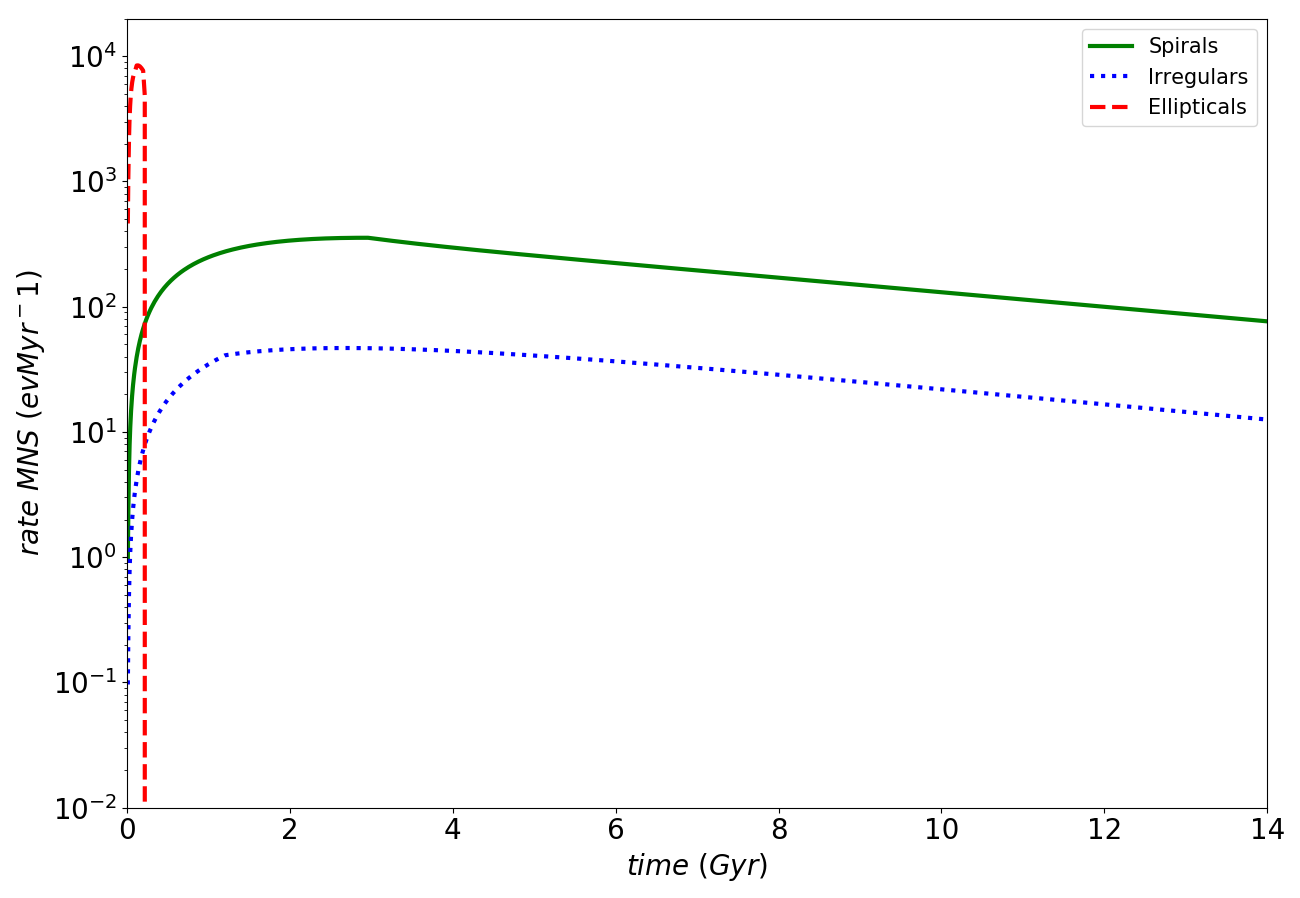}\label{fig:b}}%
 \caption{Predicted MNS rates for galaxies of different morphological type as a function of time. Panel (a): case of a DTD for MNS with $\beta=-0.9$; panel (b): case of a total delay time for all neutron stars binary systems equal to 10 Myr.}%
 \label{fig: ratedmt}%
\end{center}
\end{figure*}

In Figure \ref{fig: ratedmt} we show the predicted rates of MNS for elliptical, spiral and irregular galaxies. In particular, in panel (a) we report the results of the simulations in the case of a DTD for MNS with $\beta=-0.9$ and in panel (b) we report the results of the simulations in the case of a constant total delay time of 10 Myr. In the case of elliptical galaxies, our results show clearly which is the main effect of using a DTD instead of a constant delay time. In fact, in the case of a constant and short total delay time, the rate of MNS follows the evolution of the SFR of the simulated galaxy. Therefore, we would not expect to observe any merging event in galaxies with no SF at the present time, such as ellipticals (see Tables \ref{tab: results galaxies CC} and \ref{tab: results galaxies noCC}). On the other hand, when we hypothesize a probability distribution of delay times including long ones, the dependence of the MNS rate on the given SFR is not as strong as in the case of a constant delay time (as it is clearly expressed by equation \ref{eq:ratemns}). As a consequence, in this case the rate of MNS differs from zero in elliptical galaxies. This is a result which must be taken into account, given the fact that the host galaxy of the GW170817 event has a predominantly old population and probably no recent star formation \citep{abbott2017a}. Therefore, we conclude that a DTD including long delay times should be preferred for computing the MNS rate. In particular, among the DTDs tested here we consider as the best the one with $\beta=-0.9$, since it reproduces very well the present time observed MNS rate of \citealp{kalogera2004} for the Milky Way. 

Concerning the observed present time rate in ellipticals, there are indications (\citealp{berger, davanzo, fong}) that it should be lower ( $\sim 30\%$ of the total) but comparable with the one in spirals. Our predicted present time MNS rate in ellipticals, when assuming a DTD with long time delays, is a factor of 10 lower than the one predicted for spirals. However, our predictions cannot be really compared to those data, since they refer to a single typical galaxy for each morphological type, while the observations refer to group of galaxies with different masses and rates.

\subsection{Predicted [Eu/Fe] vs [Fe/H] pattern for spirals and the Milky Way}

\begin{figure*}
\begin{center}
 \subfloat[]{\includegraphics[width=1\columnwidth]{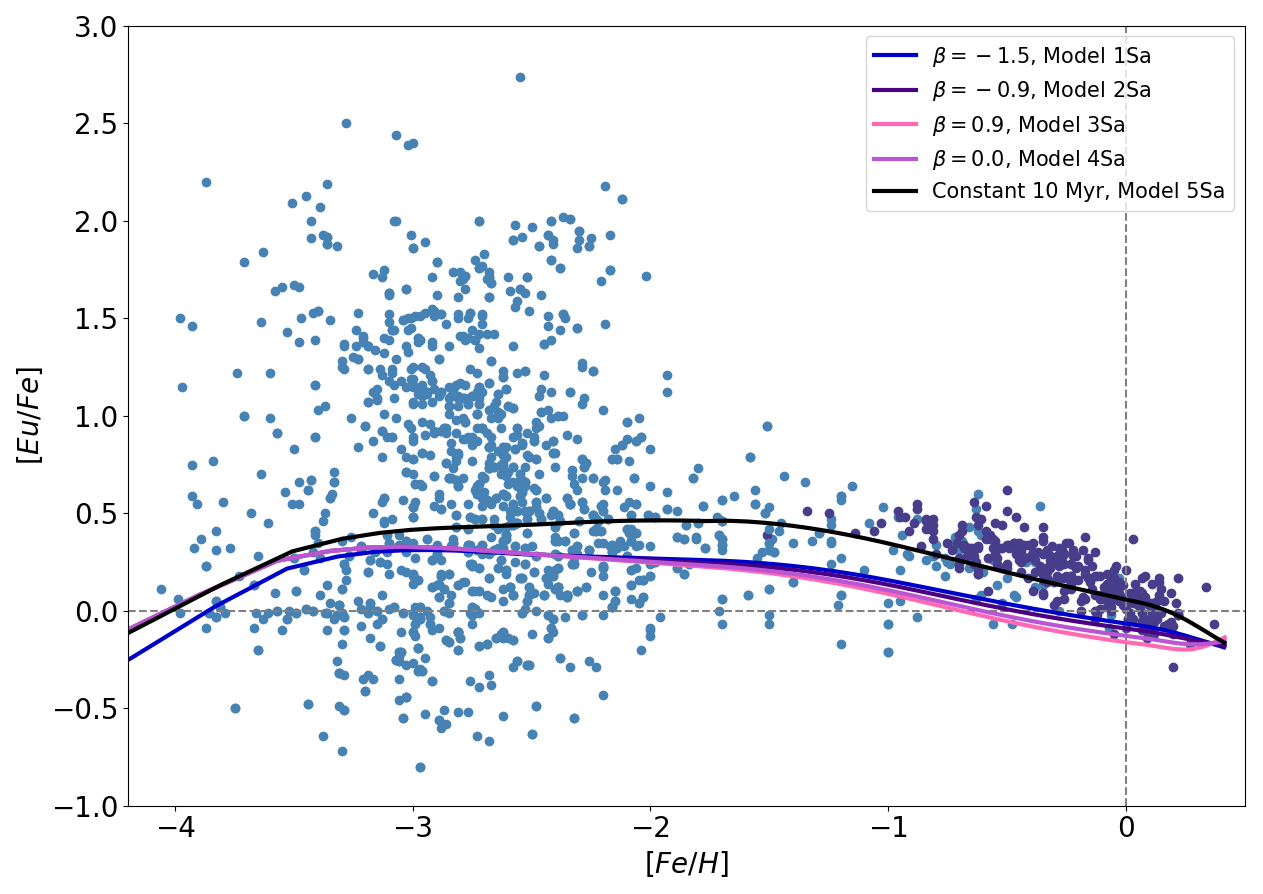}\label{fig:a}}
 \hfill
 \subfloat[]{\includegraphics[width=1\columnwidth]{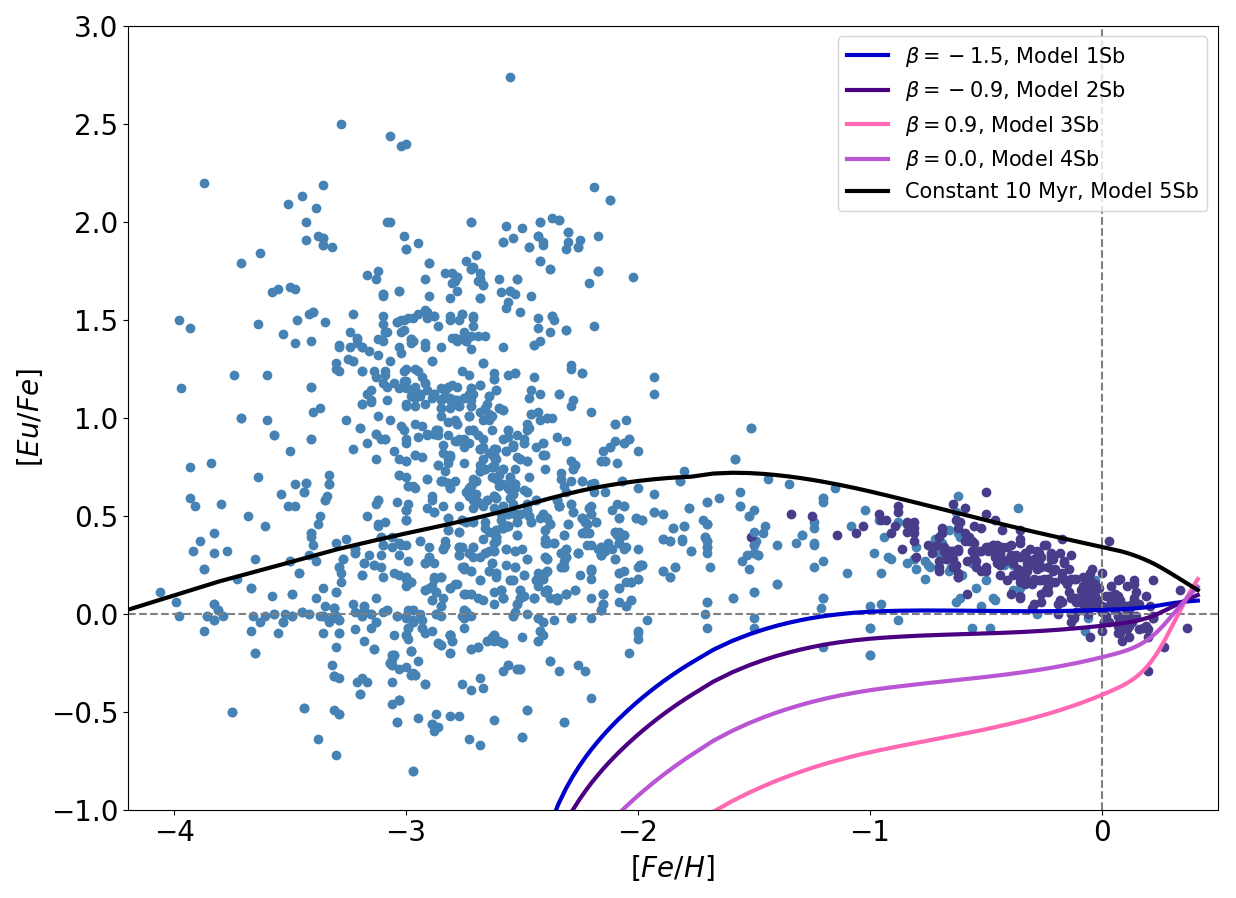}\label{fig:b}}%
 \caption{Panel (a): results of models (1-5)A with Eu production from both CC-SNe and MNS; panel (b): results of models (1-5)B with no Eu production from CC-SNe. Observational data used: $428$ Milky Way halo stars from \textit{JINA}Base and $374$ Milky Way thin disk stars from \protect\citealp{battistini}.}%
 \label{fig: EuFemodels}%
\end{center}
\end{figure*}

\begin{figure}
\begin{center}
  \includegraphics[width=1\linewidth]{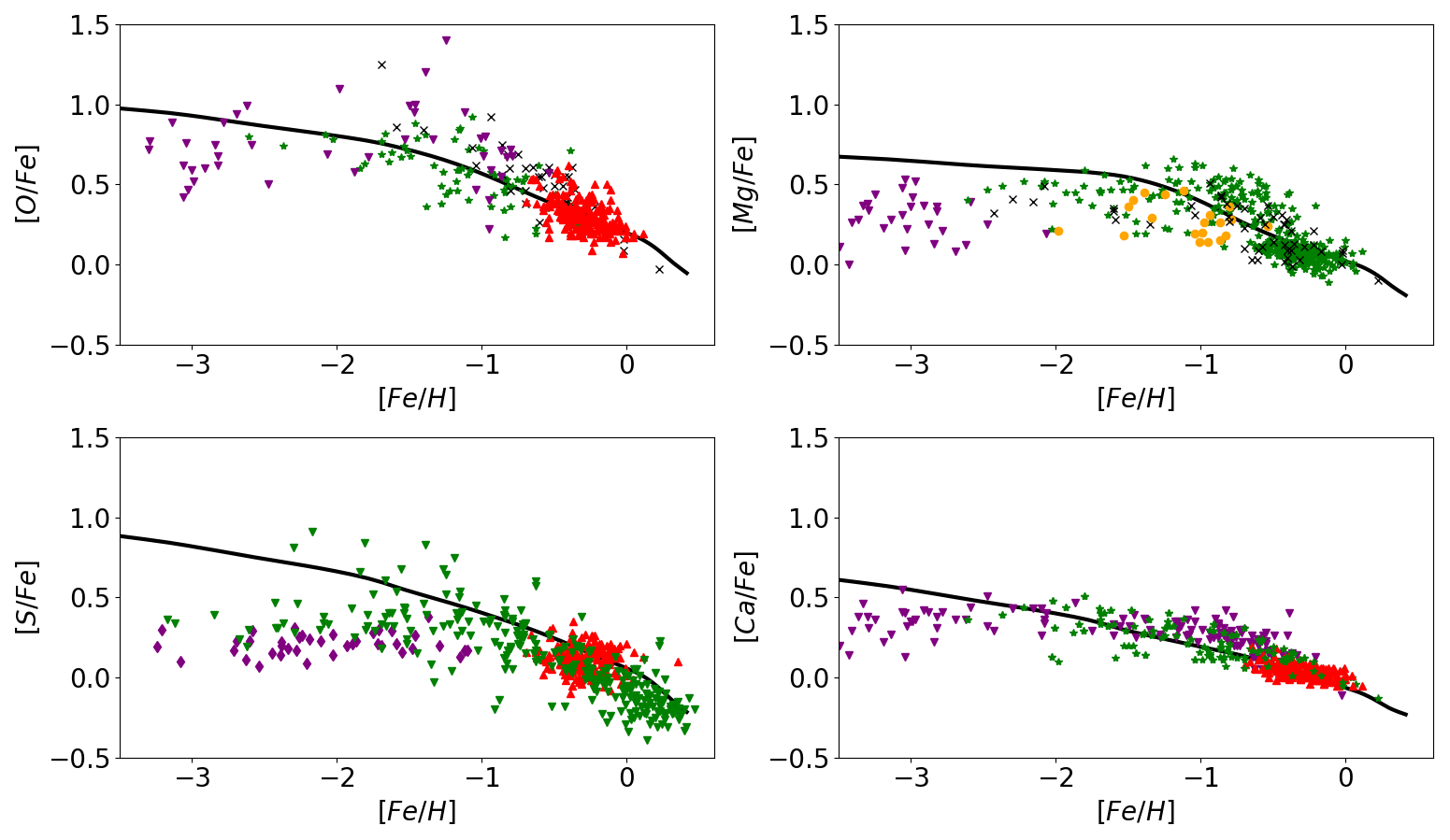}
  \caption{[$\alpha$/Fe] vs [Fe/H] predicted patterns in spiral galaxies for O, Mg, S and Ca. The observational data used are those of the Milky Way and are from \protect\citealp{cayrel04} (purple upside-down triangles), \protect\citealp{reddy03} (red triangles), \protect\citealp{gratton03} (green stars), \protect\citealp{reddy06} (orange points), \protect\citealp{ramya12} (black crosses), \protect\citealp{caffau05} (purple diamonds) and \protect\citealp{nissen07} (green upside-down triangles).}
  \label{fig: alphaMW}
 \end{center}
\end{figure}

For what concerns our predictions for the [Eu/Fe] vs [Fe/H] patterns, for spiral galaxies we tested the ability of our models to reproduce the evolution of the abundance of Eu in the Milky Way. In both panels of Figure \ref{fig: EuFemodels}, we report the observed [Eu/Fe] vs [Fe/H] relation for the Galaxy as well as the results of our simulations. For what concerns the observational data, it is possible to see that there is a large spread in the [Eu/Fe] ratio at low [Fe/H], which decreases with increasing metallicity, becoming nearly negligible for [Fe/H]$\ge -2.0$ dex. This spread in the data has been analysed by several authors (e.g. \citealp{cescutti15}; \citealp{wehmeyer}), which interpreted it as due to an initial inhomogeneous mixing. We remind that our goal here is to reproduce the main trend in the data and not the spread. The main [Eu/Fe] pattern in the Milky Way is similar to that of a typical $\alpha$-element, with a plateau in the halo phase and a decrease of the [Eu/Fe] ratio for [Fe/H]$\ge-1.0$ dex, due to the fact that for $[Fe/H]\ge -1.0$ dex, SNeIa start contributing in a substantial way to the Fe enrichment. We have therefore verified the ability of our model to reproduce the expected behaviours of four $\alpha$-elements (O, Mg, S, Ca). Those are reported in Figure \ref{fig: alphaMW}, where it is possible to see a quite good agreement with the observational data. We also remind that, since the decreasing trend at higher metallicities originates from the extra production of Fe by SNeIa \citep{matteucci09}, the production of Eu should occur on timescales shorter than SNeIa.

Coming back to Figure \ref{fig: EuFemodels}, in panel (a) we show the results of the models $(1-5)$Sa in which we consider the Eu production from CC-SNe, while in panel (b) we show the results of the models $(1-5)$Sb where MNS are the only producers of Eu (as reported in Tables \ref{tab: results galaxies CC} and \ref{tab: results galaxies noCC}).

In the case in which CC-SNe are not allowed to produce Eu, the DTDs derived by S19 are not able to reproduce neither the decreasing trend in the [Eu/Fe] for [Fe/H]$\ge -1.0$ dex, nor the plateau at lower metallicities, under-producing the [Eu/Fe] over the entire range of [Fe/H]. However, if we drop the assumption of Eu produced only by MNS and we include CC-SNe as Eu producers, we can see from the upper panel of Figure \ref{fig: EuFemodels} that the agreement with the data is improved, making us able to reproduce both the plateau in the halo phase and the decreasing trend for [Fe/H]$\ge -1.0$ dex. However, in this case the models $(1-4)$A, which differ between themselves only for the assumed DTD for MNS, do not show  much differences in their results. This is due to the fact that we have tuned the parameter $\alpha_{MNS}$ and the yield of Eu from MNS, in order to reproduce the rate of MNS and the solar abundance of Eu at the same time. With those prescriptions, CC-SNe appear to be the major producers of Eu in the simulated galaxy. For what concerns the knee at [Fe/H]$\simeq -3.5$ dex, the interpretation is that only for [Fe/H]$>-3.5 dex$ stars with $M\le 23 M_\odot$ start to die and for such stars we assumed higher yields of Eu.

In the same Figure \ref{fig: EuFemodels} models $5$Sa and $5$Sb are also shown, where all binary neutron stars systems are supposed to merge on a fixed timescale of 10 Myr. With this assumption, we are able to reproduce the expected pattern of [Eu/Fe] either in the case of Eu produced only by MNS or in the case in which Eu is produced by both MNS and CC-SNe. In particular, the model $5$Sb (with no Eu from CC-SNe) produces a higher track with respect to that produced by the model $5$Sa (with Eu from CC-SNe). This is again due to the fact that when CC-SNe do not produce Eu, we are forced to assume a higher yield of Eu from MNS (keeping fixed $\alpha_{MNS}$), in order to reproduce the observed solar abundance. It can also be observed that the model $5$B does not reproduce the flat trend of the [Eu/Fe] for [Fe/H]$< -1.0$ dex, producing instead an increasing relation. This is probably due to the fact that, while for MNS we are assuming a constant total delay time of 10 Myr, for SNeIa (the major Fe producers) we are assuming a DTD. Moreover, the nuclear lifetimes for the progenitors of MNS are assumed to be much shorter than those of the progenitors of SNeIa, which range from 40 Myr to a Hubble time.

\begin{figure}
\begin{center}
  \includegraphics[width=1\linewidth]{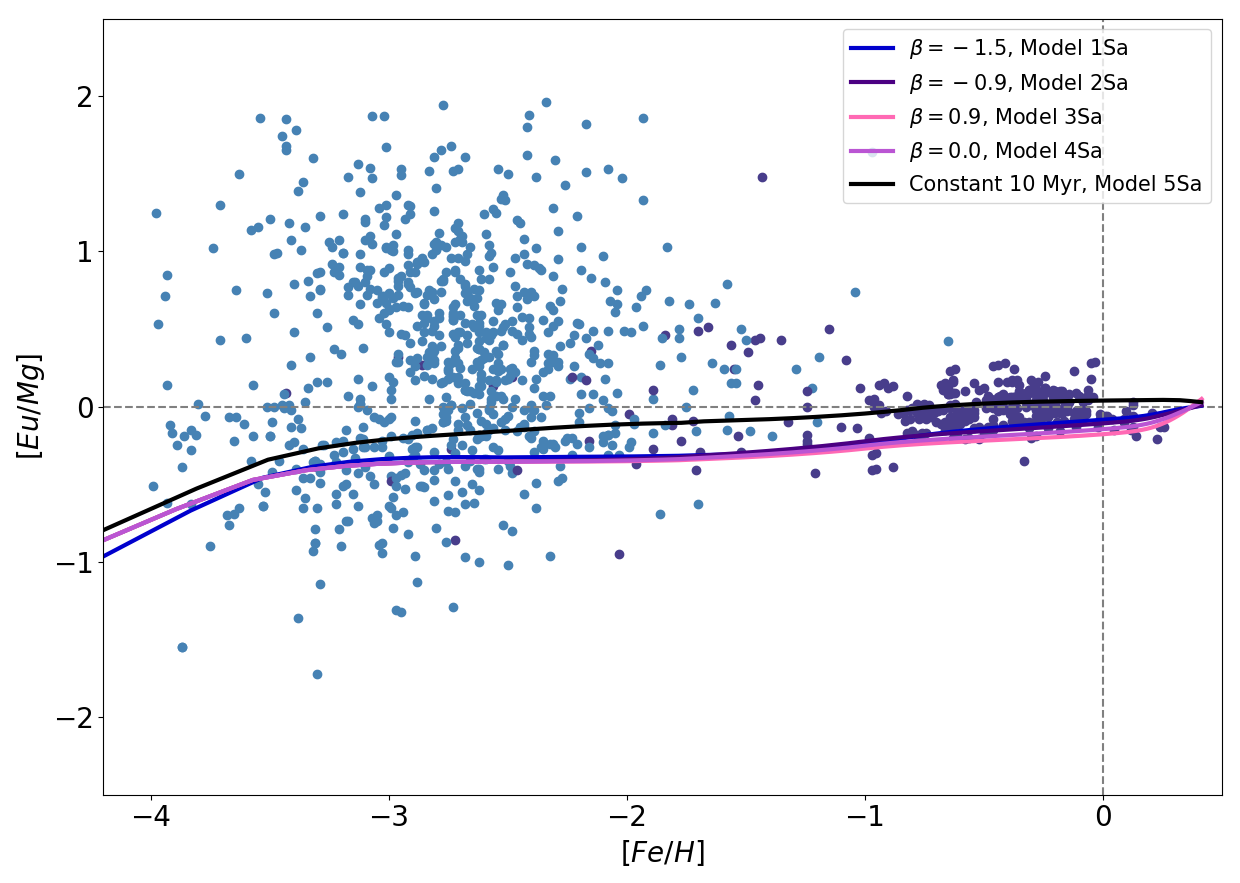}
  \caption{[Eu/Mg] vs [Fe/H] as reproduced by models (1-5)A. Observational data are a collection of $412$ Milky Way halo stars from \textit{JINA}Base and $216$ thin disk stars from a \protect\citealp{venn}).}
  \label{fig: EuMg2}
 \end{center}
\end{figure}

The fact that Fe is mainly produced by SNeIa can complicate the interpretation of the abundance ratio of [Eu/Fe] vs [Fe/H]. As suggested by \citealp{skuladottir}, it could be useful to study also the [Eu/Mg] vs [Fe/H], since Mg is a really good tracer of CC-SNe. In Figure \ref{fig: EuMg2} are reported the observed abundances of [Eu/Mg] vs [Fe/H] in the Milky Way, together with results of models ($1-5$)Sa. It is possible to see a flat trend of [Eu/Mg] with [Fe/H] data, with an average [Eu/Mg]$\simeq 0$ at all [Fe/H] and with an increasing scatter toward the lowest metallicities. If there was a significant delay in the production of Eu with respect to that of Mg, we would expect an increasing trend of the [Eu/Mg] vs [Fe/H]. The absence of such a trend in the observational data of the [Eu/Mg] for the Galaxy, suggests that the timescales of Eu and Mg production are quite similar. As we can see from the same Figure, models ($1-5$)Sa are able to reproduce the flat trend. Even in this case the model which best fits the data is the model 5Sa. However, this should be not surprising, since for this model we are assuming a fixed short merging timescale for all neutron stars systems, therefore it should not be expected a delay in the enrichment of Eu with respect to Mg. On the other hand, models ($1-4$)Sa for which a delay in the coalescence time is assumed, also show a flat trend for all [Fe/H], but with a slight increase toward high metallicities, as expected.

In conclusion, it is possible to state that when MNS are the only producers of Eu, the model which best reproduces the [Eu/Fe] vs [Fe/H] pattern in the Galaxy is the one with a constant and short total delay time of 10 Myr,  equal for all neutron star binary systems. In this case, the yield of Eu should be equal to $2.0 \times 10^{-6} M_\odot$ per merging event. On the other hand, if a DTD including longer timescales is assumed, the [Eu/Fe] pattern can be reproduced only if CC-SNe are included as Eu producers. In this case the yield of Eu per merging event is reduced to $0.5 \times 10^{-6} M_\odot$. Our values for the yields of Eu are slightly lower than those estimated from the kilonova AT$2017$gfo (which are in the range of $(3-15)\times10^{-6}M_\odot$), but they are well inside the theoretical range of $(10^{-7}-10^{-5})M_\odot$ predicted by \citealp{Korobkin}.The best DTD in the case with MNS and CC-SNe as Eu producers is again that with $\beta=-0.9$, therefore we will adopt only this DTD since now on.

\subsection{Predicted [Eu/Fe] vs [Fe/H] patterns for ellipticals and irregulars}

In section \ref{section: MNSrate} we showed that we are not able to observe any merging event in elliptical galaxies if a constant delay time of 10 Myr is assumed. Since the event GW$170817$ has been observed to come from an early-type galaxy, here we will show our predictions of the [Eu/Fe] vs [Fe/H] patterns in galaxies of different morphological type excluding the case of a constant delay of 10 Myr.

\begin{figure*}
\begin{center}
 \subfloat[]{\includegraphics[width=1\columnwidth]{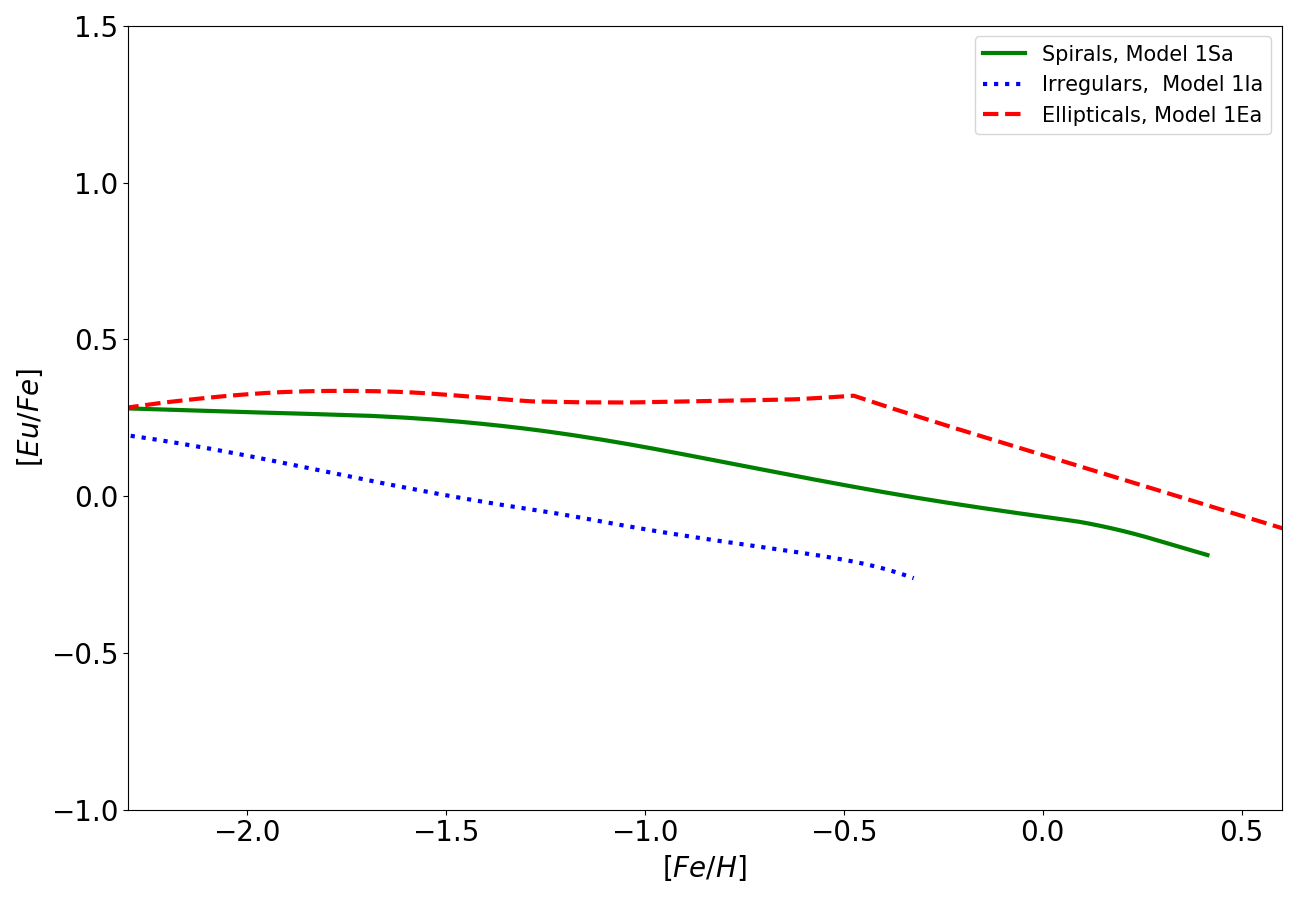}\label{fig:a}}
 \hfill
 \subfloat[]{\includegraphics[width=1\columnwidth]{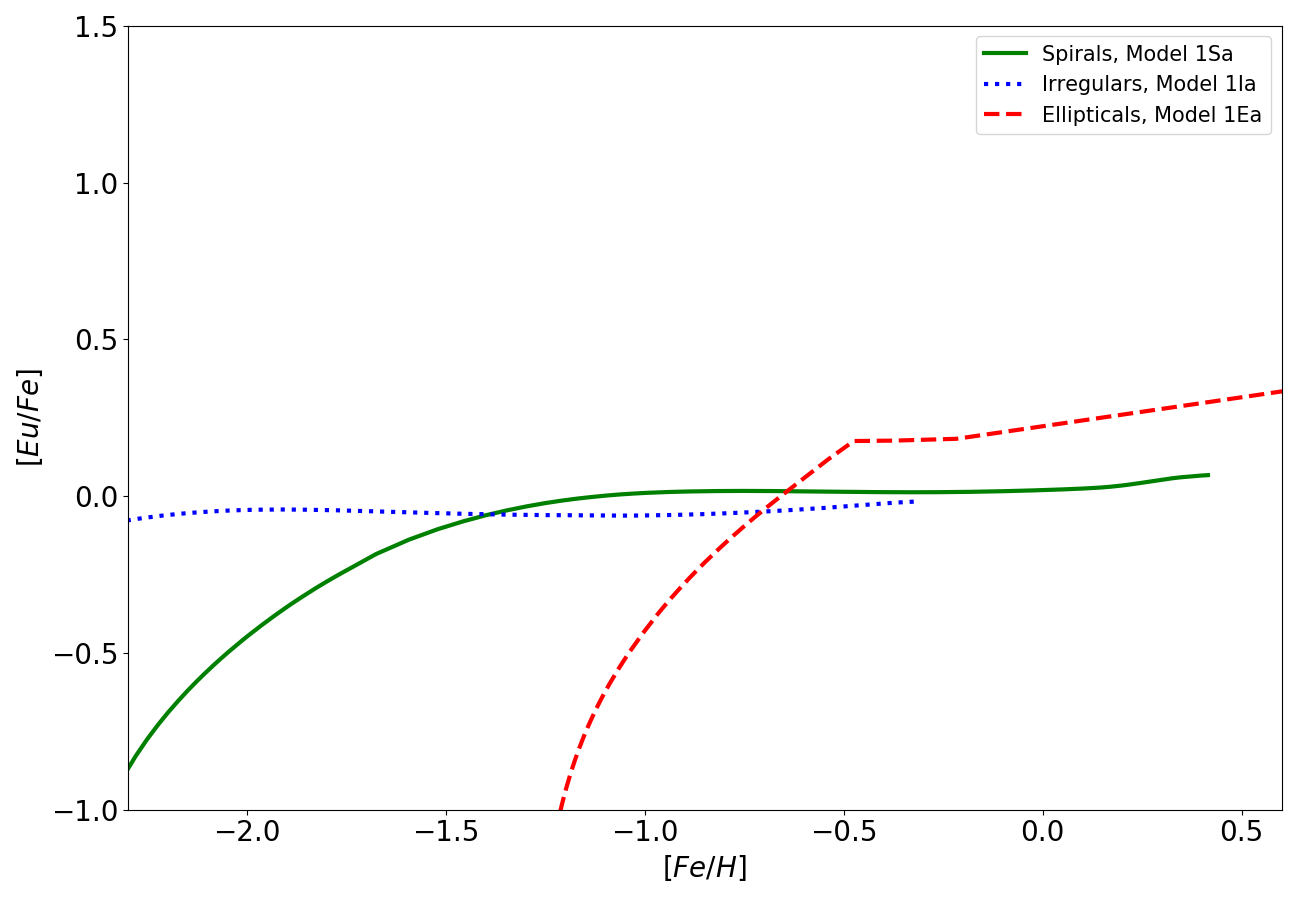}\label{fig:b}}%
 \caption{Predicted [Eu/Fe] vs. [Fe/H] by models 1S(a-b), 1I(a-b) and 1E(a-b) in which a DTD with $\beta=-0.9$ has been assumed. Panel (a): results of models with Eu from CC-SNe; panel (b): results of models with no Eu from CC-SNe.}%
 \label{fig: EuFedmtCC}%
\end{center}
\end{figure*}

In particular, in Figure \ref{fig: EuFedmtCC} we show the results of models 1S(a-b), 1I(a-b) and 1E(a-b), for which a DTD with $\beta=-0.9$ has been adopted. In panel (a) are shown the results of the models with Eu production from CC-SNe and in panel (b) are shown the results of the models for which MNS are assumed to be the only producers of Eu. 

\begin{figure}
\begin{center}
  \includegraphics[width=1\linewidth]{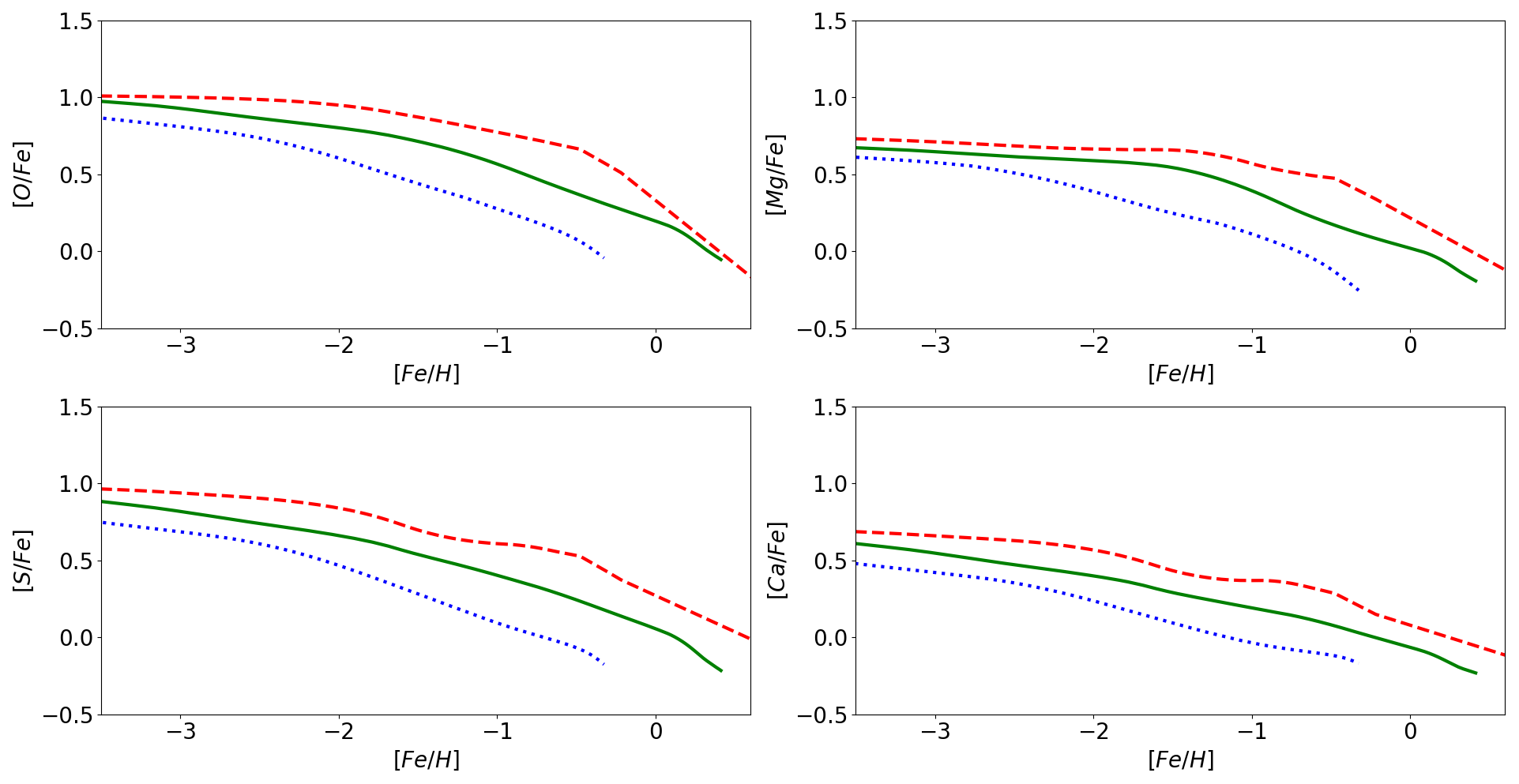}
  \caption{Predicted [$\alpha$/Fe] vs [Fe/H] patterns in galaxies with different SF histories for O, Mg, S and Ca for the same models of Figure 9.}
  \label{fig: alphadmt}
 \end{center}
\end{figure}

As one can see, our predictions are those expected on the basis of the time-delay model, according to which we would expect to find a larger plateau of the [Eu/Fe] ratio at low metallicities for ellipticals than for spirals and irregulars (see \citealp{matteucci2012}). The same behaviour holds for the $\alpha$-elements, as shown in Figure \ref{fig: alphadmt} where the predicted [$\alpha$/Fe] vs [Fe/H] patterns are reported. As one can see, for galaxies with an intense SF (ellipticals), the Fe abundance grows more rapidly because of SNeII (they also produce part of Fe). Therefore, when SNeIa (the main Fe producers) appear and the ratio [$\alpha$/Fe] begins to decrease, the Fe abundance is higher than in galaxies with a lower SF (spirals). As a consequence, the [$\alpha$/Fe] plateau extends for a larger range of metallicity (this result was first pointed out by \citealp{matteuccibrocato}). On the other hand, when the SF proceeds slowly, the Fe abundance grows less rapidly and it will be lower with respect to spirals. Therefore, for spirals and irregulars galaxies (which have an even lower SF), the [$\alpha$/Fe] plateau extends for a smaller range of [Fe/H]. Our predicted trend for the [Eu/Fe] vs [Fe/H] seems also to be in agreement with those obtained by \citealp{Grisoni} for the three main Galactic components: thick disc, thin disc and bulge. In particular, they adopt different timescales of formation and star formation efficiencies in the three Galactic component similarly to what we assume for our simulated elliptical, spiral and irregular galaxies.

\section{Cosmic Rates}
\label{cosmic}

In this section we present our analysis of the evolution of the cosmic rate of MNS (CMNSR). We define a cosmic rate as the rate in a comoving unitary volume of the Universe and we compute it as the result of the contribution of galaxies of different morphological type which are weighted according to their number densities. As already done in previous studies concerning the cosmic star formation rate (CSFR) \citep{GioanniniMatteucciCalura, caluramatteucci, vincoletto} and the cosmic SN rates \citealp{grieco}, here we will assume the galaxy number density to be a function of redshift. In particular, if $n_k$ is the number density of galaxies of the $k$-th morphological type, its evolution with the redshift $z$ can be written in the following way:

\begin{equation}
    n_k = n_{k,0}(1+z)^{\gamma_{k}},
\label{eq: nd}
\end{equation}

where $n_{k,0}$ is the number density at $z=0$ and $\gamma_k$ is the parameter which determines the evolution of the number density.\\
We consider the following cosmological scenarios:

\begin{enumerate}
    \item A pure luminosity evolution scenario (PLE), which consists of the case of $\gamma_k=0$, therefore the number density of galaxies is constant and does not evolve with redshift. In other words, all galaxies started forming at the same redshift and the number density of each morphological type has been constant since then;
    \item A number density evolution scenario (DE), which consists of the case of $\gamma_k\ne 0$, therefore the number density will evolve with redshift according to equation \ref{eq: nd}, with the parameter  shown in Table \ref{tab: nd}, the same suggested by \citealp{vincoletto}. This scenario is a typical hierarchical clustering scheme for galaxy formation, where spirals form first and then ellipticals form by subsequent mergers of spirals. In the first column we specify the morphological type of the galaxy, in the second column the number density at $z=0$ and in the third column the parameter $\gamma_k$. These parameters were chosen in order to reproduce the present time number densities of galaxies, as in \citealp{marzke};
    \item An alternative observationally based scenario, as suggested by \citealp{pozzi} and adopted by \citealp{GioanniniMatteucciCalura},  where the number density of spiral galaxies increases from $z=0$ to $z=2.3$ according to equation \ref{eq: nd}, and decreases exponentially for higher redshifts as
    \begin{equation}
        n_S = n_{0,S}(1+z)e^{-(1+z)/2}.
    \end{equation}
    In this context, ellipticals are assumed to start forming at $z=5$ and half of them form in the range $1\le z \le 2$.
\end{enumerate}

\begin{table}
\hspace{-0.5 cm}
\caption{\label{tab: nd}Parameters used for the evolution of the number density for galaxies of different morphological type. In the $1^{st}$ column it is reported the type of galaxy, in the $2^{nd}$ column the adopted number density at redshift $z=0$ and in the $3^{rd}$ column the parameter $\gamma_k$. These parameters are the same as those adopted in \protect\citealp{GioanniniMatteucciCalura}}.
\centering
\begin{tabular}{lcr}
\hline
\hline
  Type & $n_0$ $(\times 10^{-3})$ $Mpc^{-3}$ & $\gamma_{k}$\\
\hline
  Spirals & $8.4$ & $0.9$ \\
  Irregular & $0.6$ & $0.0$ \\
  Ellipticals & $2.24$ & $-2.5$ \\
\hline
\hline
\end{tabular}%
\end{table}

\subsection{Cosmic Stellar Mass Density}
\label{CSMD}

In order to compute the CMNSR, we first verify the ability of our model to reproduce the cosmic stellar mass density (CSMD). We define the CSMD as

\begin{equation}
    CSMD=\sum_{k}\rho_{*,k}(t)n_k,
\end{equation}

where $n_k$ is the galaxy number density for the $k$-th morphological type of galaxy defined in the previous section. The quantity $\rho_*(t)$ is the stellar mass density, namely the total mass density of long-lived stars (see also \citealp{madau}).

\begin{figure}
\begin{center}
  \includegraphics[width=1\linewidth]{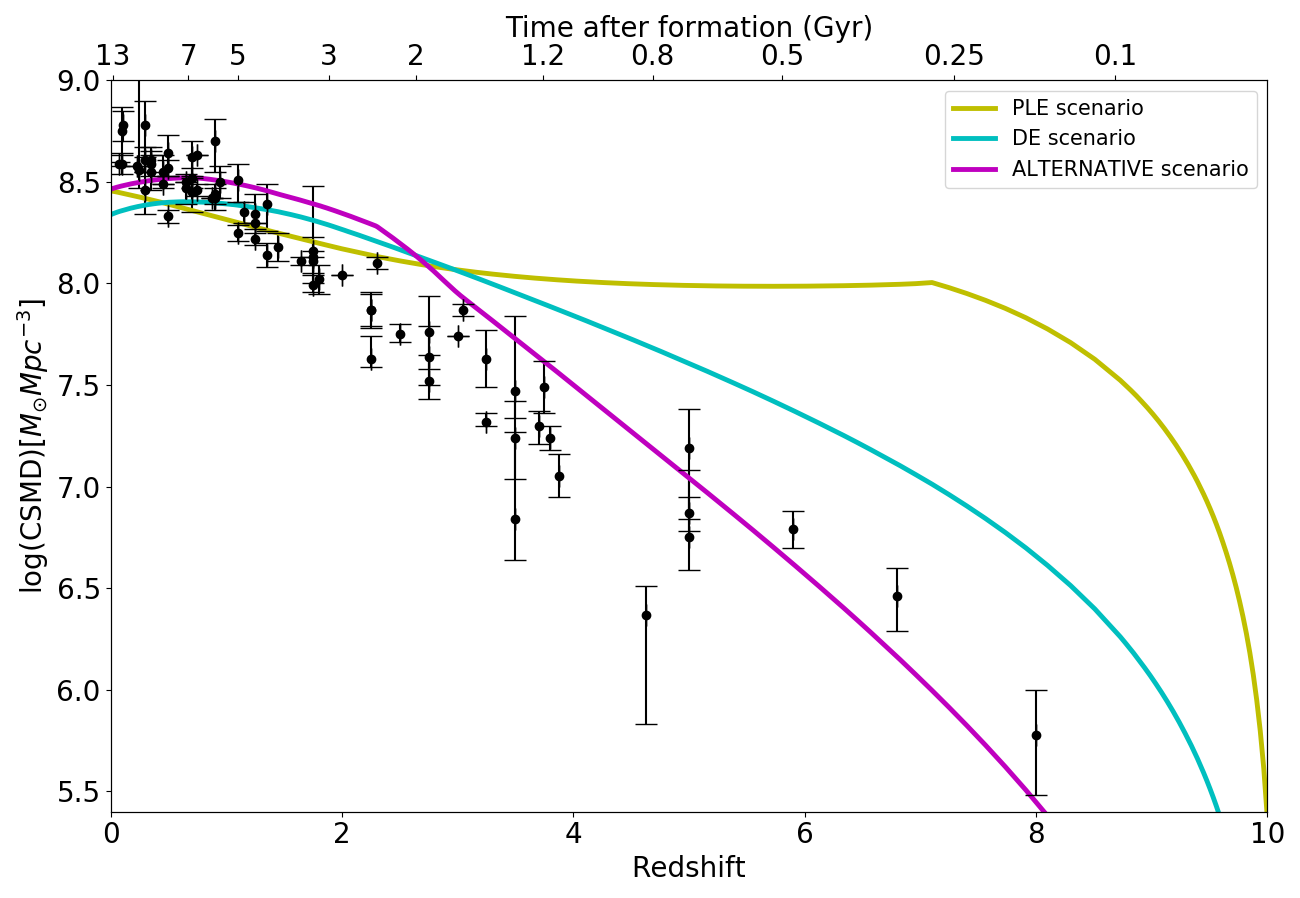}
  \caption{Cosmic stellar mass density as a function of redshift for the three different cosmological scenario of galaxy formation. Observational data are a collection from \protect\citealp{madau}.}
  \label{fig: CSMD}
 \end{center}
\end{figure}

We have computed this quantity in details by means of our galaxy models: our results for the three different cosmological scenarios of galaxy formation are shown in Figure \ref{fig: CSMD}, together with data from \citealp{madau}.

In both the DE and the alternative scenario the CSMD increases regularly over all the range of redshift, with the only difference that is steeper in the case of the alternative scenario.

On the other hand, in the PLE scenario the CSMD is characterized by a rapid increase from redshift $z=10$ to redshift $z\simeq7.5$ caused by elliptical galaxies, followed by an almost constant growth until $z\simeq2$ and a second slight increase until present time. Similar patterns have also been found for the cosmic dust mass density evolution, as shown in \citealp{GioanniniMatteucciCalura}.

All the three different scenarios are in good agreement with data at low redshift ($0 \le z \le 1$). However, the DE and in particular the PLE scenarios both predict an evolution which is too high for $z\ge2$. On the other hand, the trend in the data seems to be quite well reproduced by the alternative scenario.

By the way a comparison with the CSFR shows also that the alternative scenario is the best, as shown in \citealp{GioanniniMatteucciCalura}.

\subsection{Cosmic Merging Neutron Stars Rate}

We define the CMNSR as:

\begin{equation}
    CMNSR=\sum_{k}R_{MNS, k}(t)n_k,
\end{equation}

where $R_{MNS, k}$ is the rate of MNS at the time $t$ as defined by equation \ref{eq:ratemns} for the $k$-th morphological type of galaxy. We compute the CMNSR in the three different cosmological scenarios, both in the case of a DTD with $\beta=-0.9$ and in the case of a constant total delay time of 10 Myr. 

\begin{figure}
\begin{center}
 \subfloat[]{\includegraphics[width=1\linewidth]{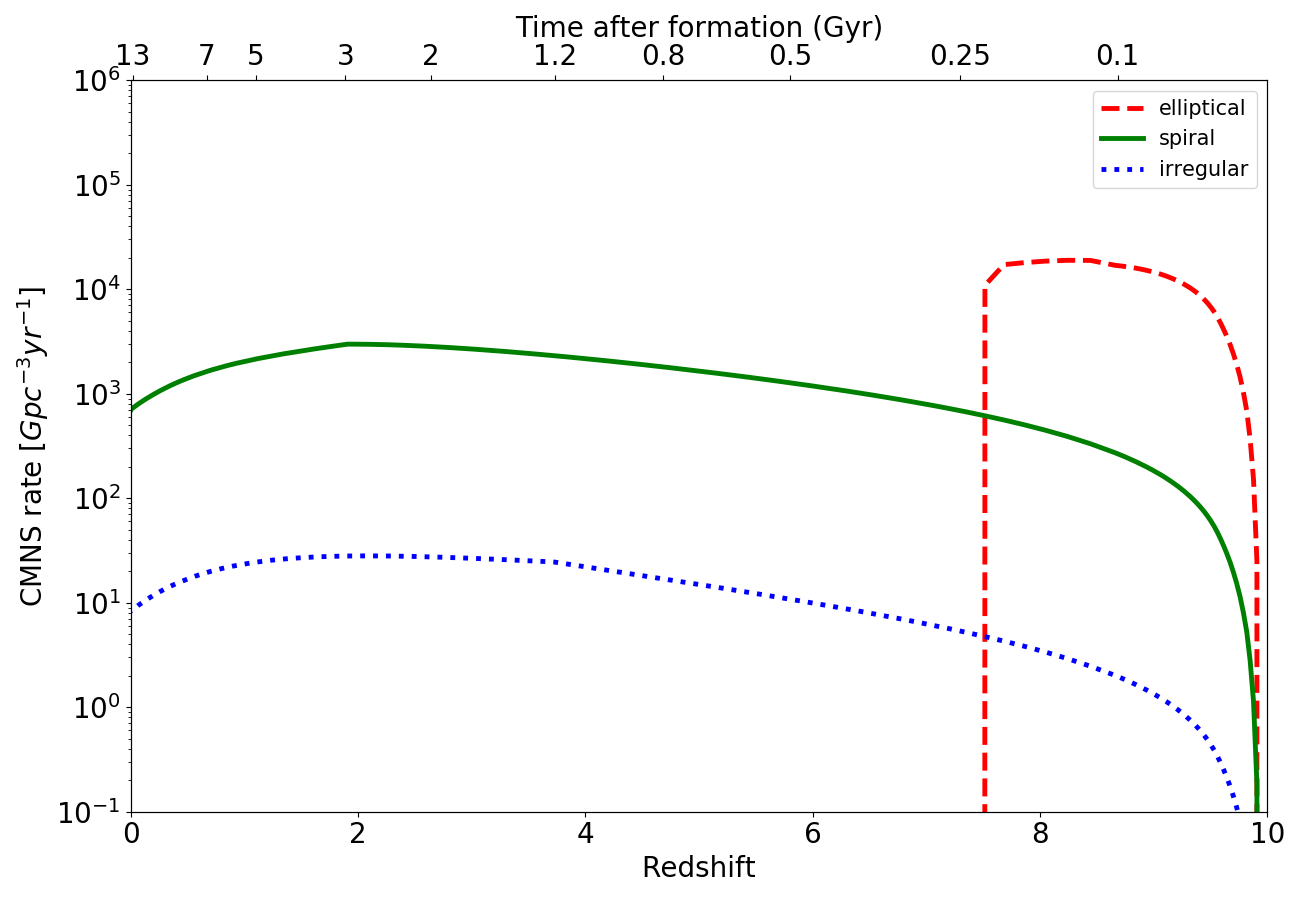}\label{fig:a}}\\
 \subfloat[]{\includegraphics[width=1\linewidth]{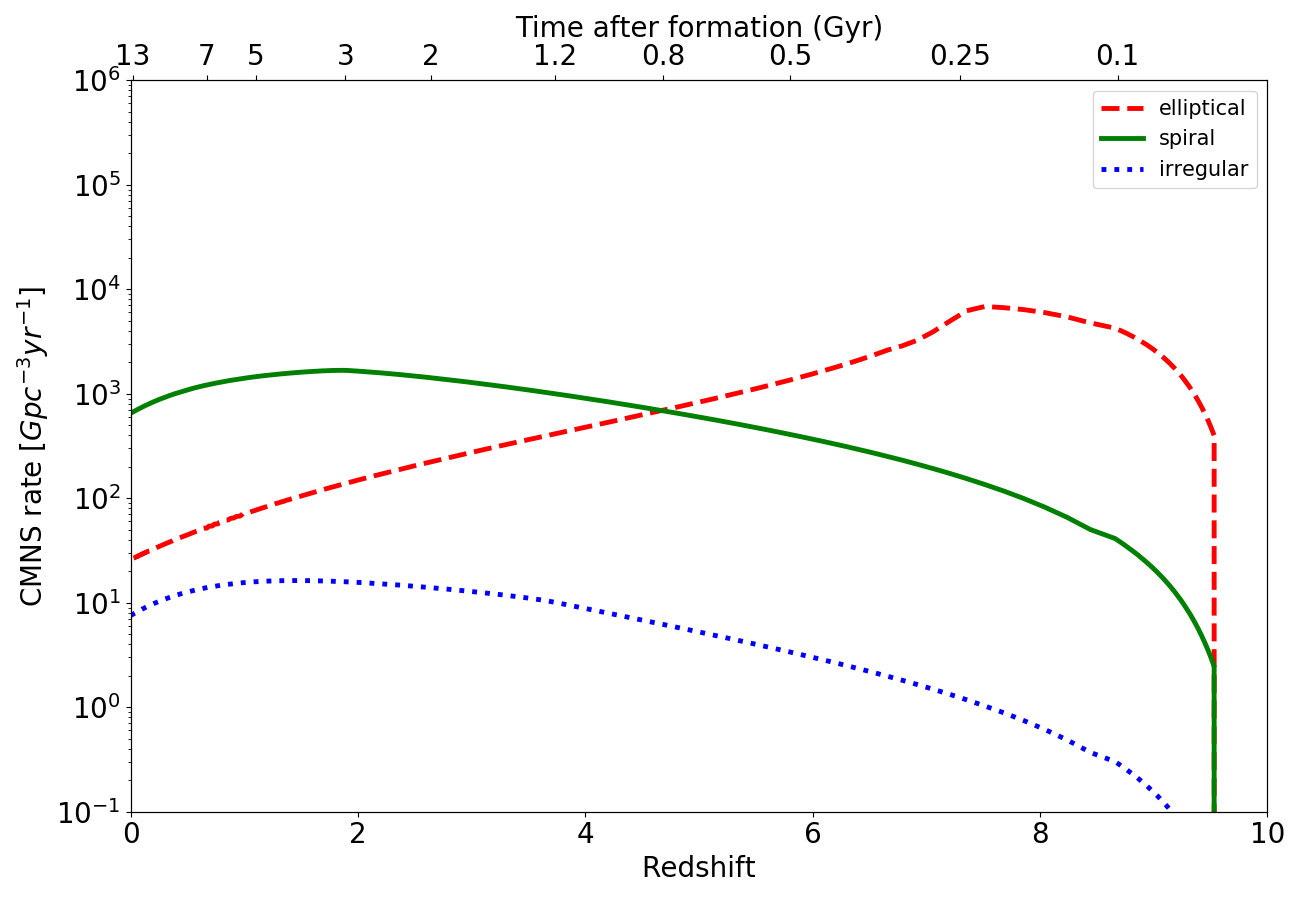}\label{fig:b}}%
 \caption{Contributions to the CMNSR from galaxies of different morphological type for the PLE scenario, in the case of a total delay time of 10 Myr (panel (a)) and in the case of a DTD with $\beta=-0.9$ (panel (b)).}%
 \label{fig: CMNSRPLE}%
\end{center}
\end{figure}

In Figure \ref{fig: CMNSRPLE} it is reported the behaviour of the CMNSR for the PLE scenario for galaxies of different morphological type, both in the case of a total delay time of $10$ $Myr$ (panel (a)) and in the one of a DTD with $\beta=-0.9$ (panel (b)). In the case of a constant total delay time, since the rate of MNS is essentially given by the integral of the SFR, we expect no contribution to the CMNSR from elliptical galaxies from redshift $z\simeq7.5$ to the present day. On the other hand, when we adopt the DTD, the elliptical galaxies contribute to the CMNSR for the whole redshift range.

\begin{figure}
\begin{center}
 \subfloat[]{\includegraphics[width=1\linewidth]{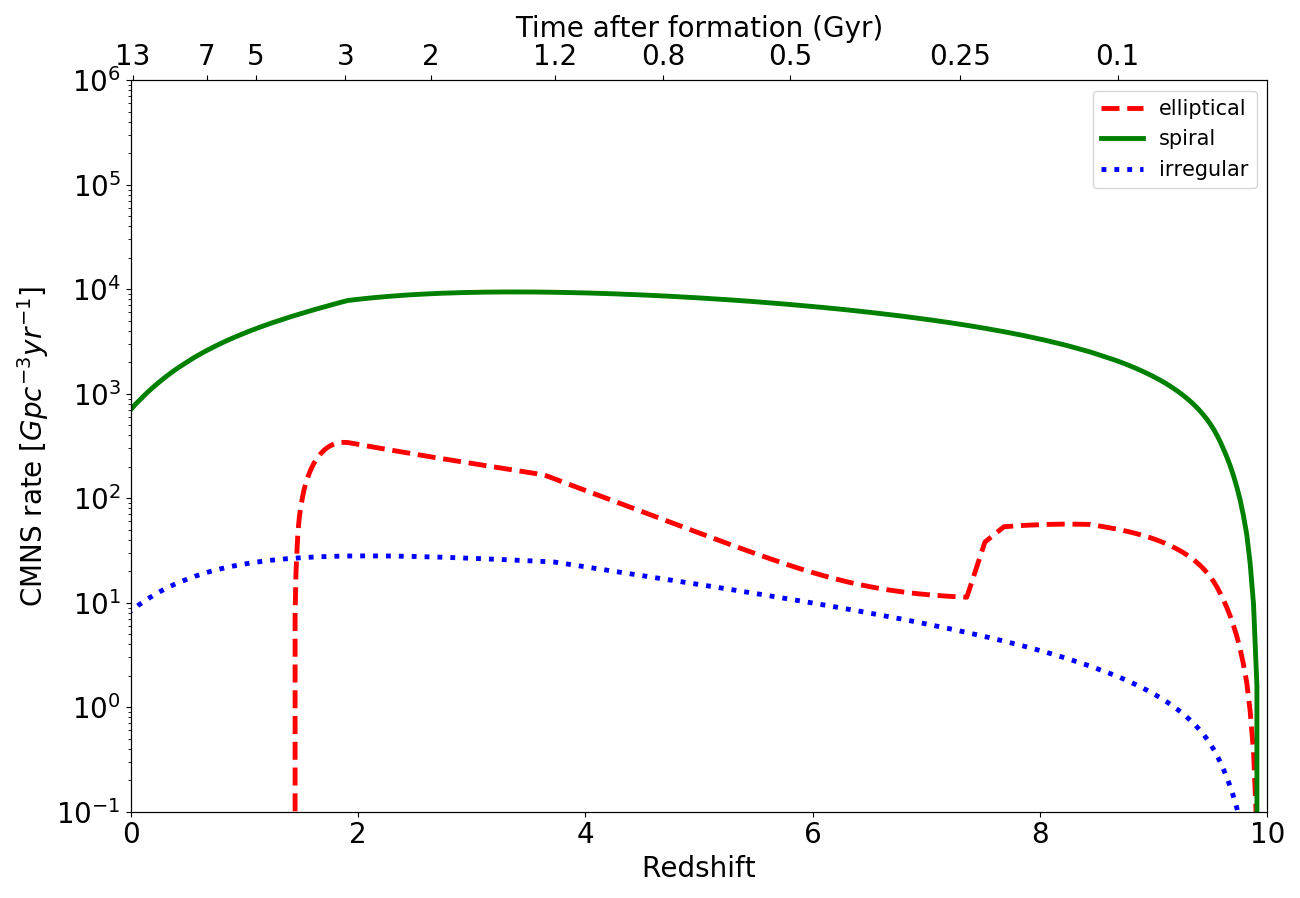}\label{fig:a}}\\
 \subfloat[]{\includegraphics[width=1\linewidth]{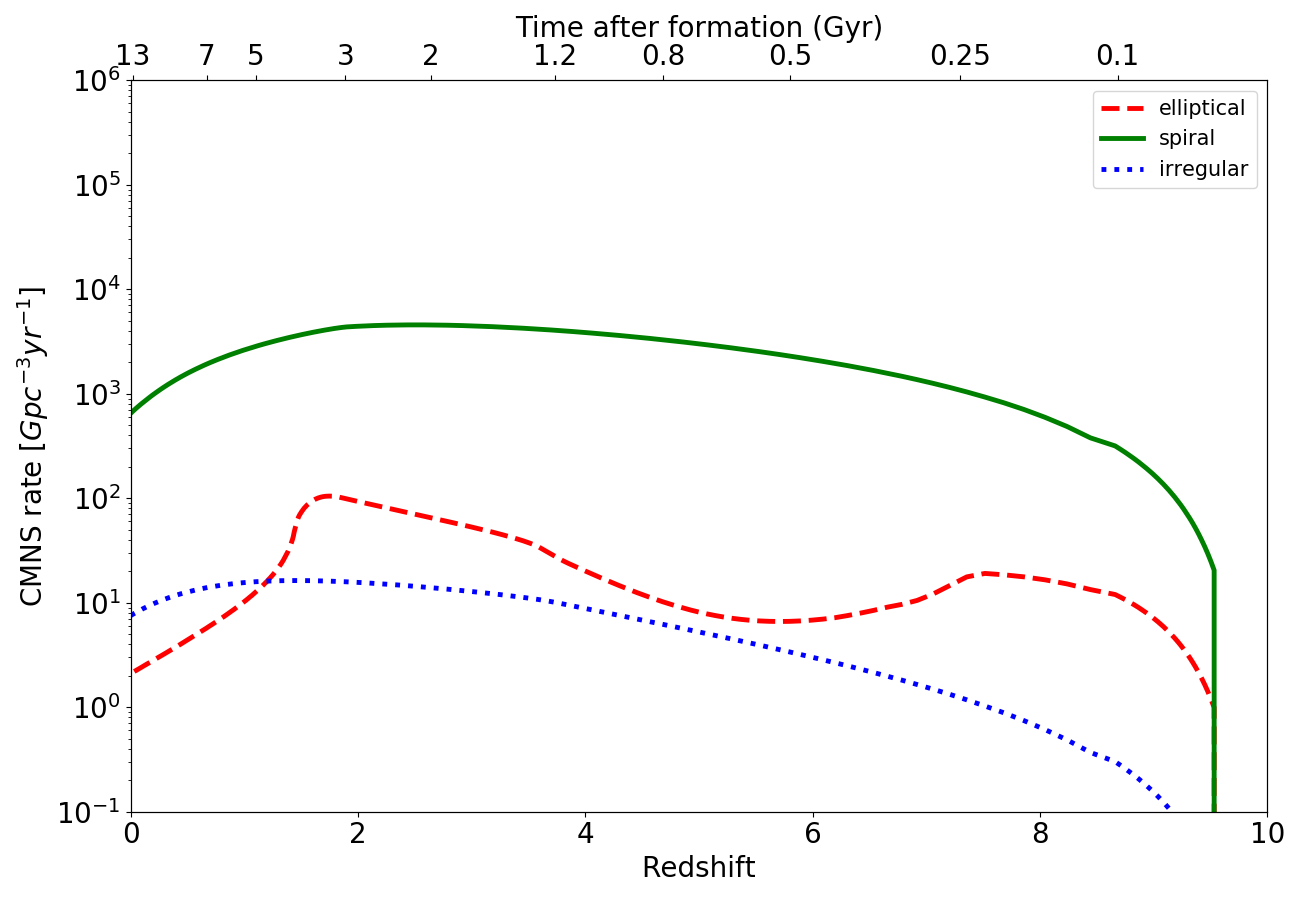}\label{fig:b}}%
 \caption{Contributions to the CMNSR from galaxies of the different morphological type for the DE scenario, in the case of a total delay time of 10 Myr (panel (a)) and in the case of a DTD with $\beta=-0.9$ (panel (b)).}%
 \label{fig: CMNSRDE}%
\end{center}
\end{figure}

\begin{figure}
\begin{center}
 \subfloat[]{\includegraphics[width=1\linewidth]{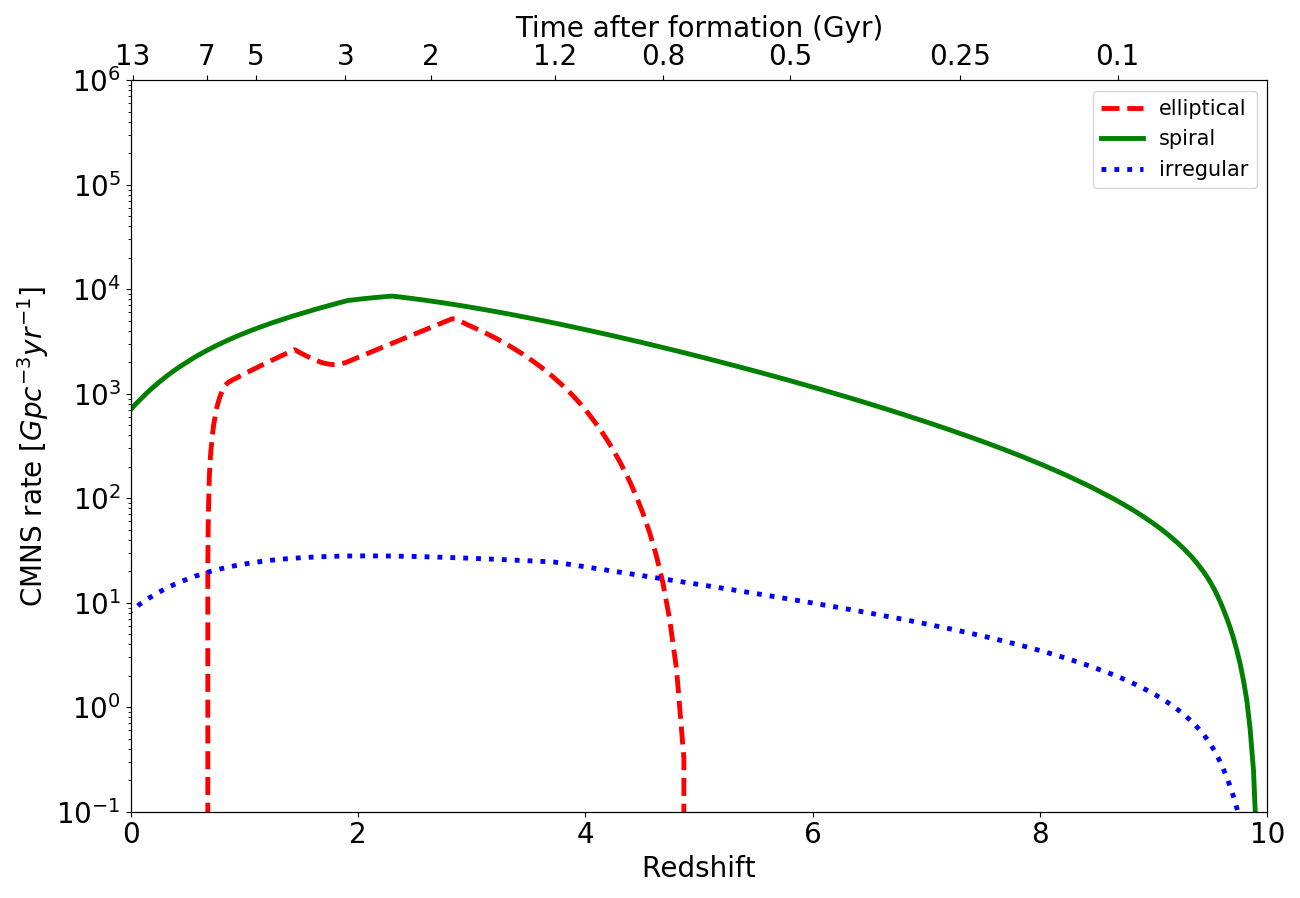}\label{fig:a}}\\
 \subfloat[]{\includegraphics[width=1\linewidth]{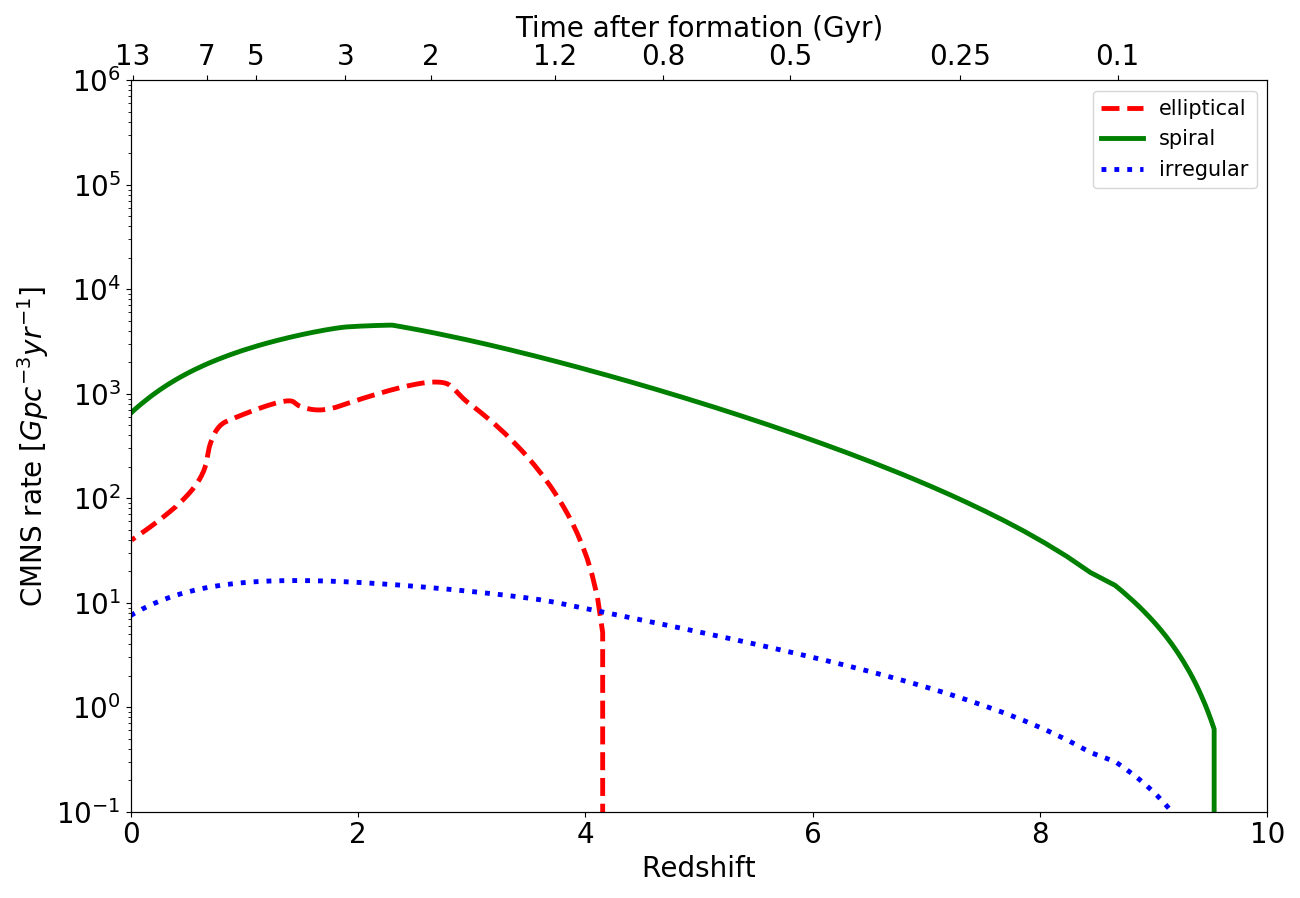}\label{fig:b}}%
 \caption{Contributions to the CMNSR from galaxies of different morphological type for the alternative scenario, in the case of a total delay time of 10 Myr (panel (a)) and in the case of a DTD with $\beta=-0.9$ (panel (b)).}%
 \label{fig: CMNSRalt}%
\end{center}
\end{figure}

In Figures \ref{fig: CMNSRDE} and \ref{fig: CMNSRalt} are shown the evolution of the CMNSR for the DE and for the alternative scenario, respectively, for galaxies of different morphological type, either in the case of a total delay time of 10 Myr (panel (a)) and in the one of a DTD with $\beta=-0.9$ which represents the best choice for this parameter, as we have seen in Section \ref{section: MNSrate}, (panel (b)). Also in these scenarios, it appears evident the effect of adopting a DTD rather than a constant time delay. However, it must be noted that in both these two scenarios the contribution to the CMNSR from elliptical galaxies has a lower impact than that of spirals. Therefore, the effect of using a DTD or a constant time delay will have almost no consequences on the total CMNSR behaviour. 

\begin{figure}
\begin{center}
 \subfloat[]{\includegraphics[width=1\linewidth]{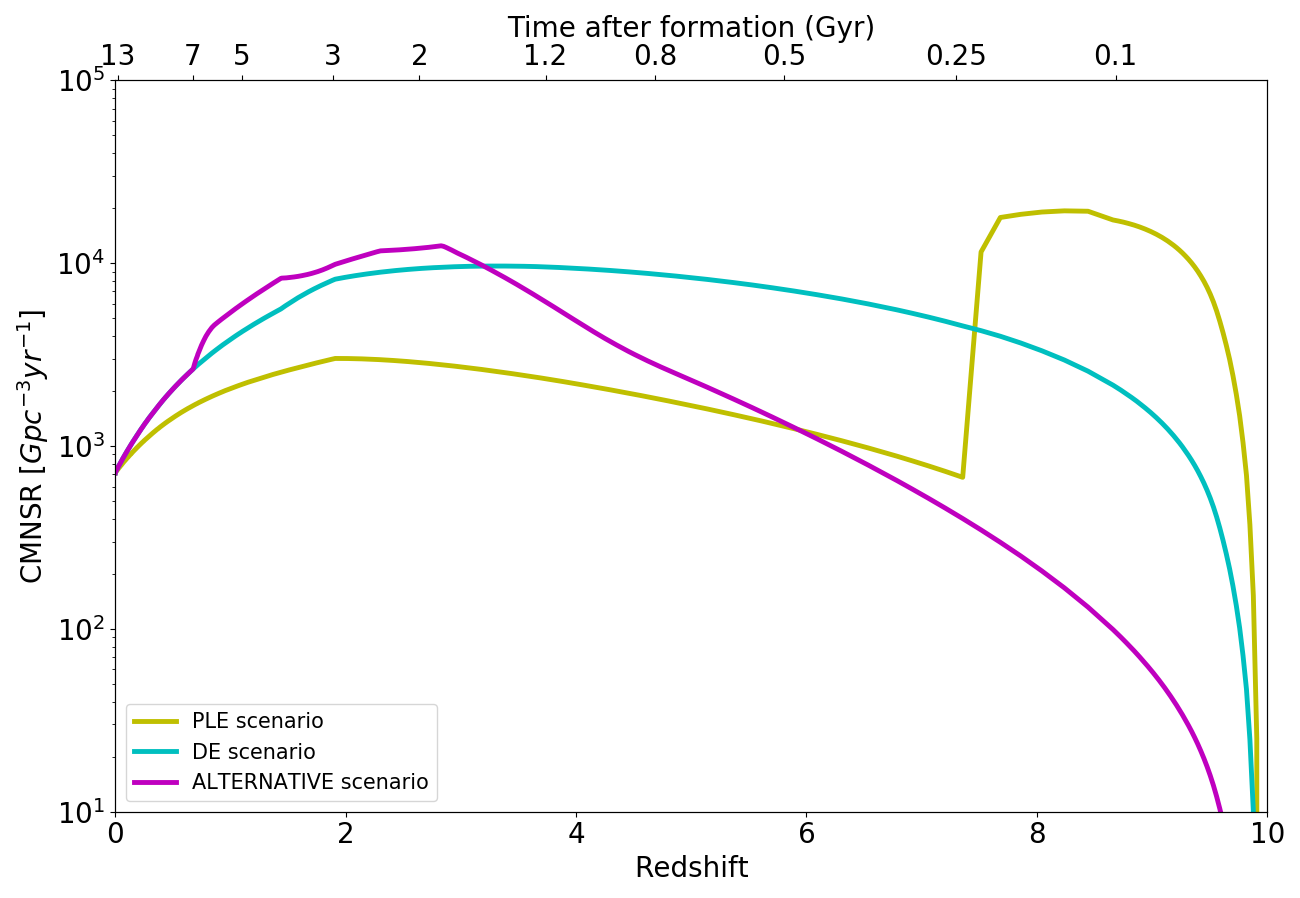}\label{fig:a}}\\
 \subfloat[]{\includegraphics[width=1\linewidth]{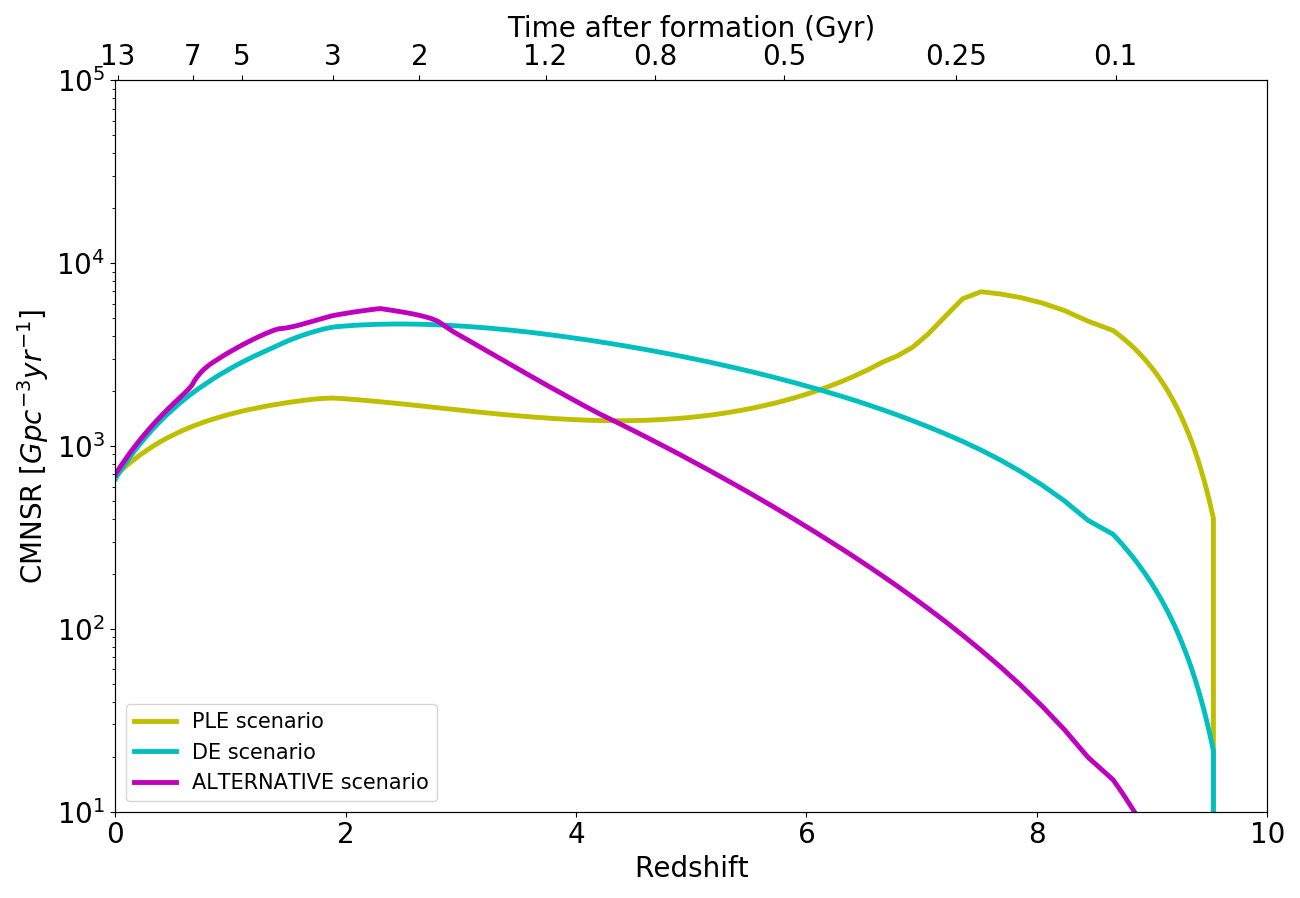}\label{fig:b}}%
 \caption{CMNSR as a function of the redshift for the three different scenarios (PLE, DE and alternative) in the case of a total delay time of $10$ $Myr$ (panel (a)) and in the case of a DTD with $\beta=-0.9$ (panel (b)).}%
 \label{fig: CMNSR}%
\end{center}
\end{figure}

Our results for the total CMNSR rate for the three different cosmological scenarios are shown in Figure \ref{fig: CMNSR}, where in panel (a) are reported the results of our simulations in the case of a constant total delay time and in panel (b) are reported the results in the case of the assumed DTD for MNS.

For what concerns the PLE scenario, in the case of a constant total delay time of 10 Myr, the CMNSR shows two peaks: the first peak is at $z\simeq8$ and a second one is at redshift $z\simeq2$. In analogy with the CSFR, the first peak is due to elliptical galaxies, which dominate the PLE scenario at highest redshift. When the star formation of elliptical galaxies stops, the CMNSR abruptly decreases and from that moment on its evolution will be due to spiral galaxies. 

In the case of a DTD with $\beta=-0.9$, the CMNSR decrease after the first peak, is smoother. As we explained this is the main effect of using a DTD.

In the DE scenario, both in the case of a constant total delay time and of a DTD, the CMNSR does not show the high redshift peak, because of the lower impact of elliptical galaxies. 

For what concerns the alternative scenario, the CMNSR is dominated by spirals for all the range of redshifts, with a peak around $z=3$. 

Finally in Table \ref{tab: CMNSR} both the total CMNSR and the single contributions to the CMNSR from galaxies of different morphological type, in the three different scenarios, are reported. In particular, in the first column we specify the DTD for MNS, in the second column the total CMNSR for the PLE, DE and alternative scenario, in the third column it is reported the contribution to the total CMNSR from spiral galaxies, in the fourth the contribution from ellipticals galaxies and in the last column that from irregular galaxies. By the end of the Table there are also reported the observed MNS rate. Our predictions for the total CMNSR in all the three different cosmological scenarios, both in the case of a constant total delay time and in the case of a DTD with $\beta=-0.9$, are consistent, within the error bars, with the rate of MNS predicted by LIGO/Virgo and equal to $1540^{+3200}_{-1220}Gpc^{-3}yr^{-1}$. Moreover, the rate of MNS in the Galaxy estimated by \citealp{kalogera2004} and equal to $\sim 80^{+200}_{-60} Myr^{-1}$ can be converted to a cosmic rate of MNS equal to $800^{+2000}_{-600}Gpc^{-3}yr^{-1}$ for a Galaxy number density of $\sim 10^{-2}Mpc^{-3}$ \citep{dellavalle}. Our predictions for the total CMNSR seem to be in a very good agreement with this last estimate.

\begin{table*}
\hspace{-0.5 cm}
\caption{\label{tab: CMNSR}CMNSR for galaxies of different morphological type and in the three different cosmological scenarios (PLE, DE and alternative). In the $1st$ column it is reported the DTD for MNS used, in the $2nd$ column it is reported the total CMNSR, in the $3rd$, $4rd$ and $5th$ columns are reported the contribution to the total CMNSR from spirals, ellipticals and irregulars galaxies, respectively. By the end of the Table we also report the observed rates of MNS as predicted by \protect\citealp{abbott2017a} and by \protect\citealp{kalogera2004}.}
\centering
\begin{tabular}{ccccc}
\hline
\hline
 & & PLE scenario & &\\
\hline
 MNS  & TOT CMNSR & Spirals CMNSR & Ellipticals CMNSR & Irregulars CMNSR \\
 DTD & $(yr^{-1}Gpc^{-3})$ & $(yr^{-1}Gpc^{-3})$ & $(yr^{-1}Gpc^{-3})$ & $(yr^{-1}Gpc^{-3})$ \\
\hline
  Constant $10$ Myr & $651$ & $643$ & $0$ & $8$ \\
  $\beta=-0.9$ & $647$ & $607$ & $33$ & $7$ \\
\hline
\hline
 & & DE scenario & &\\
\hline 
 MNS  & TOT CMNSR & Spirals CMNSR & Ellipticals CMNSR & Irregulars CMNSR \\
 DTD & $(yr^{-1}Gpc^{-3})$ & $(yr^{-1}Gpc^{-3})$ & $(yr^{-1}Gpc^{-3})$ & $(yr^{-1}Gpc^{-3})$ \\
\hline
  Constant $10$ Myr & $751$ & $706$ & $0$ & $8$ \\
  $\beta=-0.9$ & $662$ & $651$ & $3$ & $7$ \\
\hline
\hline
 & & ALTERNATIVE scenario & &\\
\hline 
 MNS  & TOT CMNSR & Spirals CMNSR & Ellipticals CMNSR & Irregulars CMNSR \\
 DTD & $(yr^{-1}Gpc^{-3})$ & $(yr^{-1}Gpc^{-3})$ & $(yr^{-1}Gpc^{-3})$ & $(yr^{-1}Gpc^{-3})$ \\
\hline
  Constant $10$ Myr & $715$ & $706$ & $0$ & $8$ \\
  $\beta=-0.9$ & $711$ & $651$ & $52$ & $7$ \\
\hline
\end{tabular}%

\begin{tabular}{ccc}
\hline
\hline
 & MNS observed rates $(yr^{-1}Gpc^{-3})$ &\\
\hline
 Central Value & Confidence Interval & Reference \\
\hline
 $800$ & $200-2800$ & \citealp{kalogera2004} \\
 $1540$ & $320-4740$ & \citealp{abbott2017a} \\
\hline
\end{tabular}
\end{table*}

\subsection{Comparison with the SGRB redshift distribution}

\begin{figure}
\begin{center}
 \subfloat[]{\includegraphics[width=1\linewidth]{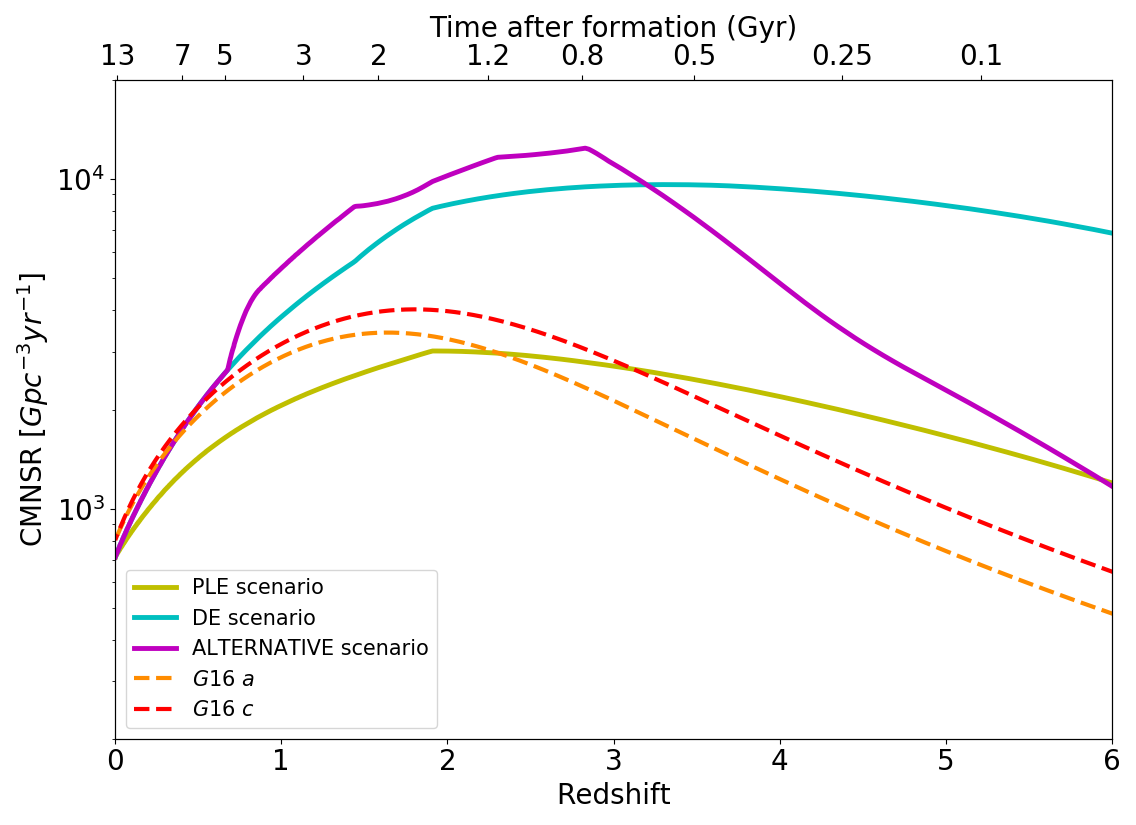}\label{fig:a}}\\
 \subfloat[]{\includegraphics[width=1\linewidth]{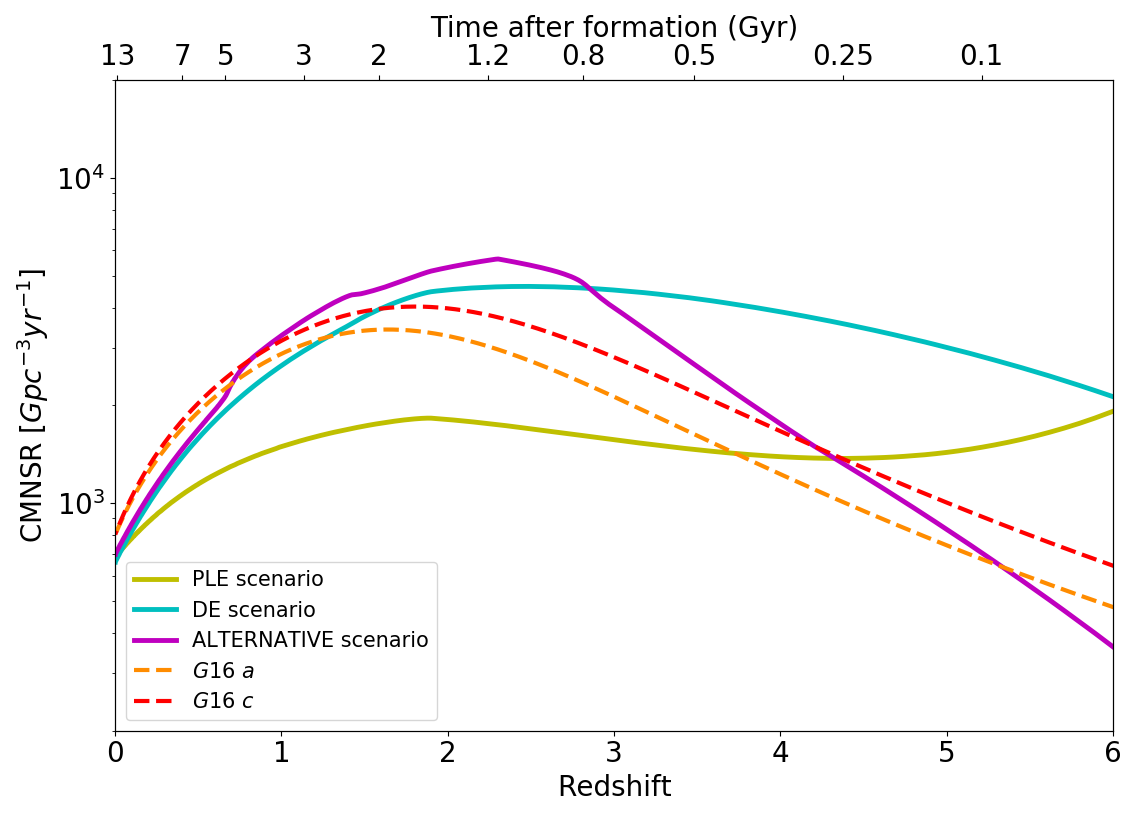}\label{fig:b}}%
 \caption{Redshift distributions of SGRB as found by \citealp{G16} (their case 'a', orange dashed line; their case 'c', red dashed line) against our predicted redshift distribution of MNS rate in the three different cosmological scenarios (PLE, DE and alternative), in the case of assuming a total delay time of 10 Myr (panel (a)) and in the case of a DTD for MNS with $\beta=-0.9$ (panel (b)).}%
 \label{fig: sgrbcmnsr}%
\end{center}
\end{figure}

Our predicted redshift distributions of the total CMNSR is directly comparable with the redshift distributions of SGRBs derived by \citealp{G16}. To make the comparison possible, the SGRBs redshift distribution has been multiplied by a suitable factor in order to reproduce the cosmic rate of MNS. The comparison can be seen in Figure \ref{fig: sgrbcmnsr}, for the three cosmological scenarios and both in the case of a total delay time of $10 Myr$ (panel (a)) and in the case of DTD for MNS with $\beta=-0.9$ (panel (b)). We remind that our cosmic rate has been obtained by weighting the rate of MNS in galaxies of different morphological type. We do not compare our predicted rate of MNS in a single galaxy with the observed SGRBs rate. In fact, we model only one typical galaxy per morphological type, while observations refers to galaxy of the same type but different masses. 

We will consider the alternative scenario as the best one, since it also best reproduces both the CSMD (\ref{CSMD}) and the cosmic star formation rate (CSFR), as shown in previous studies \citep{GioanniniMatteucciCalura, grieco}. The PLE scenario, in fact, is probably unrealistic at high redshift (due to the fact that in this scenario we are neglecting the evolution of the number density) and it also underestimates data at low redshift. The DE scenario, on the other hand, has a smoother evolution, but it is shown to overestimate data at low redshift.

With that in mind, our results seem to be in good agreement with those previously found by S19. In other words, the rate of SGRBs as proposed by \citealp{G16} is best represented by a bottom heavy distribution, which here is given by a DTD for MNS with $\beta=-0.9$. This value of $\beta$ gives rise to a distribution of time delays which scales basically as $\propto t^{-1}$. On the other hand, in the case of a total delay time of $10 Myr$ we are not able to obtain a good agreement with \citealp{G16} SGRBs distributions, since we are producing a maximum which seems to overestimate those predicted by \citealp{G16} by a significant factor. 

\citealp{G16} find an isotropic rate of SGRBs equal to $0.2$  $yr^{-1}Gpc^{-3}$ (their model (a)). After comparison \textit{a posteriori} their derived rate with the rate of MNS derived from population synthesis models and from the statistics of Galactic binaries, they infer an average opening angle $\theta$ of $3^{\circ} - 6^{\circ}$. From this value, we can derive a beaming correction factor\footnote{$f_b^{-1} \equiv (1-cos\theta)^{-1}$, where $\theta$ is the jet opening angle.} of $(183-730)$, so that the true rate will be $(36-146) yr^{-1}Gpc^{-3}$. We can compare this value with the one obtained by \citealp{wandermanpiran}, which find an isotropic rate of SGRBs of $4.1$ $yr^{-1}Gpc^{-3}$ and suggest that a typical beaming correction factor would be $30$ (as also suggested by \citealp{fong2012}). By correcting their observed rate by this factor, we find a true rate of $123$ $yr^{-1}Gpc^{-3}$, comparable with that derived by \citealp{G16}.

\subsection{Predicted number of Kilonovae detections for future surveys}

On the basis of our results for the contributions to the CMNSR from galaxies of different morphological type (Figures \ref{fig: CMNSRPLE}, \ref{fig: CMNSRDE} and \ref{fig: CMNSRalt}), we can predict the number of Kilonovae detections for future LSST and VST surveys. The sky areas and the corresponding volumes patrolled by LSST and VST for Kilonovae detections have been taken from \citealp{dellavalle} and are equal to $\simeq 2.1 Gpc^{3}$ and $\simeq 0.13 Gpc^{3}$, respectively. The predicted number of Kilonovae detections in different morphological type of galaxies and for the three cosmological scenarios are reported in Tables \ref{tab: kn1} and \ref{tab: kn2}. In particular, in Table \ref{tab: kn1} we report the predicted rates obtained hypothesizing a DTD with $\beta=-0.9$ and in Table $\ref{tab: kn2}$ are reported those computed adopting a constant delay of $10$ Myr. In both Tables we report in the first column the morphological type of galaxies, in the second column the cosmological scenario and in the third and fourth columns the predicted rate for LSST and VST, respectively. For spiral and irregular galaxies we report only the predicted detections in the PLE scenario, since there are no significant differences in the CMNSR among different cosmological scenarios.

These predictions should be corrected by a factor $\eta$ that accounts for the "efficiency" of a given survey. $\eta$ depends on several parameters, such as the control time (i.e. a quantity that depends on the survey cadence and on the photometric time scale evolution of the transient, see e.g. \citealp{1cappellaro}), sky conditions, technical downtime and scheduling constraints. Typical values are of the order of $\eta \simeq 50\%$, but they can be as low as $5\%$ \citealp{grado}.

The Transient High-Energy Sky and Early Surveyor (THESEUS) space mission \citealp{amati}, currently under study by the European Space Agency (ESA) for a possible launch in 2032, would be able to reveal electromagnetic counterparts of MNS ad SGRBs up to $z \sim 1$ \citealp{stratta}. THESEUS should detect a few dozens of SGRBs on-axis per year.

\begin{table}
\hspace{-0.5 cm}
\caption{\label{tab: kn1}  Predicted Kilonovae rates ($ yr^{-1} Gpc^{-3}$) for LSST and VST surveys in the case of a DTD with $\beta=-0.9$. In the $1^{st}$ it is specified the type of galaxy, in the $2^{nd}$ column the cosmological scenario, in the $4^{th}$ and $5^{th}$ columns the predicted Kilonovae detections for LSST and VST surveys, respectively.}
\centering
\begin{tabular}{cccc}
\hline
\hline
  Type & Scenario & LSST & VST\\
\hline
  Spiral & PLE & $1850$ & $91$ \\
  Irregular & PLE & $21$ & $1$ \\
  Elliptical & PLE & $74$ & $4$ \\
  Elliptical & DE & $17$ & $0$ \\
  Elliptical & alternative & $130$ & $6$ \\
\hline
\hline
\end{tabular}%
\end{table}

\begin{table}
\hspace{-0.5 cm}
\caption{\label{tab: kn2} Predicted Kilonovae rates ($ yr^{-1} Gpc^{-3}$) for LSST and VST surveys in the case of a constant total delay time of 10 Myr. In the $1^{st}$ it is specified the type of galaxy, in the $2^{nd}$ column the cosmological scenario, in the $4^{th}$ and $5^{th}$ columns the predicted Kilonovae detections for LSST and VST surveys, respectively.}
\centering
\begin{tabular}{cccc}
\hline
\hline
  Type & Scenario & LSST & VST\\
\hline
  Spiral & PLE & $2215$ & $100$ \\
  Irregular & PLE & $28$ & $1$ \\
  Elliptical & PLE & $0$ & $0$ \\
  Elliptical & DE & $0$ & $0$ \\
  Elliptical & alternative & $0$ & $0$ \\
\hline
\hline
\end{tabular}%
\end{table}

\section{Conclusions}
\label{conclusions}

In this paper we have computed for the first time the rate of MNS in galaxies of different morphological type (ellipticals, spirals and irregulars) and we have studied the effect of these events on the chemical evolution of a typical r-process element, namely the Eu, in the Milky Way, taken as a typical spiral galaxy.  The models for galaxies of different type differ mainly by the history of star formation, with the ellipticals suffering a strong and short burst of star formation, while the spirals and even more the irregulars suffer a continuous and more moderate star formation. The adopted galaxy models can reproduce the main features of typical galaxies of each type, as shown by \citealp{GioanniniMatteucciCalura}. The evolution of the Eu in the Milky Way has been investigated  by assuming either that MNS are the only producers of this element or that both CC-SNe and MNS contribute to the Eu enrichment.
 For the Milky Way we also tested four different delay time distributions, the same derived by S19 and corresponding to four different values of the $\beta$ parameter ($-1.5$, $-0.9$, $0.0$, $0.9$), as well as a constant total delay time of 10 Myr equal for all neutron star binary systems. In particular, we have tuned the parameters of our simulations in order to reproduce the following observational constraints of the Milky Way: the present time rate of MNS, the solar Eu and Fe abundances and the [Eu/Fe] vs. [Fe/H] relation.\\
Our main results for spiral galaxies can be then summarized as follows:

\begin{enumerate}
    \item The present time rate of MNS in the Galaxy is well reproduced either by assuming a DTD with parameter $\beta=-0.9$ (DTD $\propto t^{-1}$)  or a constant total delay time of 10 Myr. In the first case, the occurrence probability of MNS is $5.42\%$ while in the second case it is found to be $6.15\%$;
    \item The [Eu/Fe] vs. [Fe/H] in the Milky Way can be reproduced by assuming only MNS as Eu producers only if a constant total delay time (stellar lifetime plus gravitational time delay) of 10 Myr is adopted, a result already suggested by previous works (M14; \citealp{cescutti15}). In this case, the yield of Eu for merging event should be $2.0\times 10^{-6} M_\odot$, which is slightly lower than those estimated from the kilonova AT$2017$gfo, but well inside the theoretical range of $(10^{-7}-10^{-5})M_\odot$ predicted by \citealp{Korobkin} ;
    \item If a DTD including long time delays for MNS is assumed, the [Eu/Fe] vs. [Fe/H] in the Milky Way is reproduced only if CC-SNe are also included as Eu producers.  The best DTD has $\beta=-0.9$. In this case, the yield of Eu for merging event should be $0.5\times 10^{-6} M_\odot$, while the yield of Eu from CC-SNe is the modified version of the model SN$20150$ from \citealp{argast}. In this case, CC-SNe become the main production site of Eu, in agreement with S19 results for the Milky Way. 
\end{enumerate}

For what concerns the chemical evolution of ellipticals and irregulars, our main results can be summarized as follows:

\begin{enumerate}
    \item We are not able to see any merging event at the present time in ellipticals galaxies if a constant total delay time of 10 Myr is assumed, since in this case the rate of MNS will follow the evolution of the SFR, which has stopped $1\sim$ 10 Gyr ago in elliptical galaxies;
    \item If we instead assume a DTD ($\beta=-0.9$), namely a distribution of gravitational time delays, the present time rate of MNS is different from zero. This fact is in agreement with the probability that the galaxy NGC $4993$, host of the event GW$170817$, is an early-type galaxy with an old dominant stellar population \citealp{abbott2017a};
    \item The predicted [Eu/Fe] vs. [Fe/H] patterns in galaxies of different morphological type follow the expectations of the time-delay model, either in the case of MNS being the only Eu producers or if  CC-SNe also participate to the Eu production.This means a longer plateau for [Eu/Fe] in the case of ellipticals with higher SFR, and a shorter plateau for irregulars with a weak SFR. (see \citealp{matteucci2012}).
\end{enumerate}

We have also studied the cosmic evolution of the stellar mass density and  MNS rate in three cosmological scenarios: (i) a PLE  (pure luminosity evolution) scenario, in which the number density of the different morphological types of galaxies does not evolve with redshift; (ii) a DE  (density evolution) scenario, in which the number density is assumed to evolve with redshift in order to reproduce a typical hierarchical galaxy formation; (iii) an alternative scenario, where both spirals and ellipticals are assumed to evolve on the basis of observational constraints (see \citealp{pozzi}), which is very similar to the DE scenario. The CMNSR has been computed both in the case of a DTD with $\beta=-0.9$ and  of a constant total delay time of 10 Myr. Our results have been then compared with the redshift distributions of SGRBs, as  derived by \citealp{G16}.\\ Our conclusion are the following:

\begin{enumerate}
    \item The CSMD is best reproduced in the case of an alternative scenario of galaxy formation. By the way, this same scenario is also the best to reproduce the CSFR, as shown already by Gioannini et al. (2017).On the other hand, in both the PLE and the DE scenarios our results overestimate the data for redshift $z \ge 2$;
    \item The CMNSR in the PLE scenario, with a constant delay time for MNS, shows a first peak at redshift $z\sim 8$, due to the high redshift formation of ellipticals. When the star formation in ellipticals stops, the CMNSR abruptly decreases and its evolution is then due to spiral galaxies, leading to a second peak at redshift $z\sim 2$;
    \item If a DTD for MNS is assumed, the decrease of the CMNSR after the high redshift peak in the PLE scenario is smoother, since it does not stop with the quench of star formation. Therefore, in this case the elliptical galaxies will contribute to the CMNSR during the whole range of redshift. In particular, the contribution from ellipticals appears to be dominant at high redshift, whereas that from spirals is dominant at lower redshift;
    \item Both in the DE and  alternative scenario, the contribution to the CMNSR from elliptical galaxies has a lower impact with respect to that of spirals. Therefore, the effect of using a DTD or a constant and short time delay has almost no consequences on the total CMNSR behaviour;
    \item Our predictions of the present time CMNSR in all the three different cosmological scenarios, both in the cases of a constant total delay time and DTD, are consistent with the rate of MNS observed by LIGO/Virgo and the one estimated by \citealp{dellavalle};
    \item Assuming the alternative scenario as the best one (see also \citealp{GioanniniMatteucciCalura}), the SGRBs redshift distribution proposed by G16 is best represented by our CMNSR with an assumed  DTD $\propto t^{-1}$. On the other hand, in the case of a constant delay time too many events at higher redshift with respect to G16 are produced. Therefore, we conclude that in order to reproduce the SGRB rate, the assumption of a constant total delay time should be rejected. However, S19 found a way to reconcile a short time delay, good for reproducing the [Eu/Fe] ratio in the Galaxy, with the SGRBs rate; that is by assuming that the percentage of systems giving rise to a MNS event ($\alpha_{MNS}$) is variable in time. The adoption of a variable $\alpha_{MNS}$ will be the subject of future work.
\end{enumerate}

Finally, for what it concerns our predictions of the number of Kilonovae detections for the LSST and VST surveys, we can conclude that:
\begin{enumerate}
    \item On the basis of the results shown in Tables \ref{tab: kn1} and \ref{tab: kn2}, it is possible to state that, at least in principle, the observations of the number of MNS can be used to discriminate the different scenarios at play. However, Kilonovae are intrinsically weak objects detectable only up to $z\sim0.25$ and $z\sim0.05$ with LSST and VST, respectively. A comparison with Figures \ref{fig: CMNSRPLE}, \ref{fig: CMNSRDE} and \ref{fig: CMNSRalt} indicates that the number of MNS detections occurring in spirals and irregulars at low $z$ cannot be used to disentangle among different scenarios, since they provide similar results. Differences in results emerge only at very high values of redshift, which, unfortunately are not observable with current/future structures. On the other hand, each scenario is capable of predicting significantly different Kilonovae rates in ellipticals even at very low $z$. Therefore, observations of MNS in early-type galaxies are of the utmost importance because they can effectively help to discriminate models characterized by a constant total delay ($10$ Myr) or by delay time distribution functions.
\end{enumerate}

\section*{Acknowledgements}
M. M. aknowledges Marco Palla for some useful suggestions. P.S. and F.M. thank Laura Greggio for many useful discussions. Also, we thanks an anonymous referee for carefully reading the paper and for all the useful suggestions.

\section*{Data Availability}
The data underlying this article will be shared upon request.















\bsp	
\label{lastpage}
\end{document}